\documentclass[twocolumn,epsfig,pre,showpacs]{revtex4}

\def\be{\begin{equation}}
\def\ee{\end{equation}}
\def\bea{\begin{eqnarray}}
\def\eea{\end{eqnarray}}

\def\vs{\vec{S}}
\def\d{d\hspace{-.16cm}^{-}}
\def\vh{\vec{h}}
\usepackage{graphicx}
\usepackage{graphics}
\usepackage{amsmath}
\usepackage{hyperref}
\usepackage{latexsym}

\def\be{\begin{equation}}
\def\ee{\end{equation}}
\def\bea{\begin{eqnarray}}
\def\eea{\end{eqnarray}}

\def\vs{\vec{S}}
\def\vh{\vec{h}}
\draft

\begin{document}
\title{Nonequilibrium steady states of the isotropic classical magnet}
\date{\today}
\author{Jayajit Das$^{1}$, Madan Rao$^{2,3,\star}$
and Sriram Ramaswamy$^{4,\dagger}$}
\affiliation{
$^1$Department of Chemical Engineering and Materials Science Division,
Lawrence Berkeley National Laboratory, 
University of California, Berkeley, CA 94720, USA\\
$^2$Raman Research Institute, C.V. Raman Avenue,
Sadashivanagar, Bangalore 560080,
India\\
$^3$National Center for Biological Sciences, UAS-GKVK Campus, Bellary Road,
Bangalore 560065, India\\
$^4$ Center for Condensed Matter Theory,
Department of Physics, Indian Institute of Science, Bangalore 560012, India} 

\begin{abstract}
We drive a $d$-dimensional Heisenberg magnet using a spatially anisotropic current of mobile particles or heat. The
continuum Langevin equation is analyzed using a dynamical renormalization group, stability analysis
and numerical simulations. We discover a rich steady-state phase diagram,
including a critical point in a new nonequilibrium universality class, and a
spatiotemporally chaotic phase. The latter may be `controlled' in a robust manner
to target spatially periodic steady states with helical order. We discuss several physical
realizations of this model and make definite predictions which could be tested in experimental or model lattice systems.
\end{abstract}


\maketitle

\section{Introduction}

There is a growing awareness that spatio-temporal chaos may be a generic feature of driven,
dissipative, spatially extended systems with reversible terms in their equations of motion
\cite{EPL,UWE,UJZ}  
An understanding of this interplay between external drive,
irreversible dissipation and reversible (inertial) dynamics may even be relevant to a
deeper understanding of fluid turbulence described by the Navier Stokes equations.  In
light of this, a useful strategy is to construct simple spatially extended systems whose
dynamics reveals this interplay \cite{ZOLTAN1,ZOLTAN2} 
and study its consequences in a systematic manner. In an
earlier shorter communication \cite{EPL}, we had successfully constructed such a driven, spatially
extended system which, while ordered in the absence of the drive, inevitably 
enters a spatiotemporally chaotic state when driven.  In this longer paper, in addition to providing analytical and numerical
details of our earlier work, we present some new results and discuss several microscopic
realizations of the dynamics. We hope this study will encourage searches for 
possible realizations in lattice models and, more important, experimental systems.

The simplest examples of driven, spatially extended systems is the class of models called
driven diffusive lattice gas models (DDLG) \cite{SZIA,MARROBOOK}.  
While these models give rise to a variety of
interesting dynamics, none of them is chaotic. Recall that the DDLG
order parameter is Ising-like; its dynamics is therefore purely dissipative and externally
driven and has no reversible (inertial) terms in the equation of motion.  The simplest way
to incorporate reversibility is to elevate the Ising to Heisenberg spins; the driven
dynamics of the $O(3)$ Heisenberg spins has a natural precessional dynamics \cite{MAMAZ}.

How do we drive the Heisenberg magnet spatio-temporally chaotic ?  The dynamics of the
classical Heisenberg magnet possesses several conserved quantities (in fact there are
infinitely many in dimension $d=1$) \cite{FOGEDBY}. 
One possibility is to destroy some of these
conservation laws by driving it, not with an external field (which breaks the 
$O(3)$ symmetry
explicitly) but by imposing a background steady current of heat or particles (or any other
{\it mobile} species) which couples to the Heisenberg spins. If the dynamics of these
mobile species is `fast', then the imposed steady current may alter the effective dynamics
of the isotropic magnet in a way so as to destroy some of the conservation laws of the
un-driven system.

In fact, as we saw in \cite{EPL}, 
the classical $O(3)$ Heisenberg model in $d \equiv d_{\perp}+1$ dimensions 
\cite{MAMAZ} in the presence of  
a uniform current in one spatial direction (${\|}$), while retaining isotropy in the 
order parameter space, should be governed by the equations of motion 
\begin{eqnarray}
\frac{\partial {\vs}}{\partial t} & =  & \bigg( r_{\|}\partial_{\|}^2 
+ r_{\perp}\nabla_{\perp}^2 \bigg) {\vs} - v
{\vs}-\frac{u}{6}({\vs}\cdot{\vs}){\vs}
-\lambda {\vs}\times \partial_{\|}{\vs} \nonumber \\
 &   &  +g_{\|}{\vs}\times \partial_{\|}^2{\vs}
+g_{\perp}{\vs}\times \nabla_{\perp}^2{\vs} + \vec{\eta}\, . 
\label{eq:dyeq}
\end{eqnarray}
Eq. (\ref{eq:dyeq}) is a 
natural generalization of the DDLG models to the case of a 
3-component {\em axial}-vector order parameter ${\vs}$ and, as such, is an important step in 
the exploration of dynamic universality classes \cite{HH} far from equilibrium
\cite{UWE, MRUIZ}. 
The local molecular field in which spins precess in this driven state  
is dominated by the nonequilibrium precession term involving $\lambda$, 
responsible for all the remarkable phenomena we predict, 
including a novel nonequilibrium 
critical point and, in a certain parameter range, a type of turbulence.
Note that the noise $\vec \eta$ in our equations is a {\em non-conserving scalar} noise
even though the dynamics of the Heisenberg model is spin-conserving 
in the absence of driving. 
We shall comment below on the origin of this non-conservation as well 
as that in the deterministic terms. 

Here are our results : (i) Despite $O(3)$ invariance in the order-parameter 
space, the dynamics does not conserve magnetization; (ii) As a temperature-like 
parameter is lowered, 
the paramagnetic phase of the model approaches a nonequilibrium critical point in a 
new dynamic universality class controlled by the driving --- we evaluate the dynamic $z$,
roughening $\chi$ and anisotropy $\zeta$ exponents to leading order in $\epsilon = 4-d$; (iii) Below this
critical point, in mean-field
theory without stochastic forcing, paramagnetism, ferromagnetism and helical order are all
linearly unstable; (iv) Numerical
studies in space dimension $d=1,2$, in the absence of stochastic forcing,  
show spatiotemporal chaos in this last regime. This
chaos, when `controlled', is replaced by spatially periodic steady helical states 
which are robust against noise. 

We also discuss a variety of explicit lattice and continuum
realizations of this long length scale dynamics with the hope that this will stimulate 
a search for experimental systems, e.g., isotropic magnets carrying a 
steady particle or heat current, as well as model magnetized lattice-gas simulations, where
the predictions of our model can be tested. Our work reinforces the idea that
spatiotemporal chaos is a generic feature of
driven, dissipative, spatially extended systems with nonlinear reactive terms.

The generic occurrence of spatio-temporal chaos in this model is encouraging. It is
instructive to compare this model with the dynamics of an incompressible fluid, embodied by
the Navier-Stokes equations. Ignoring the fact that the while the total momentum of the
fluid $\rho {\bf v}$ is conserved, the total spin in our model is not, the reader should
not miss the formal similarity between the driven term ${\vs} \times \partial_{\parallel}
{\vs}$ and the nonlinear $({\bf v}\cdot {\nabla}) {\bf v}$ term in the Navier-Stokes,
despite their disparate origins. The similarity of the convective nonlinearity of
Navier-Stokes and the precession term in the dynamics of the classical Heisenberg model has
already been remarked on \cite{MAMAZ,FNS};  the form of our drive term (bilinear with {\em
one} gradient) makes this resemblance closer. Our equations are however more amenable to
quantitative analysis, since, unlike Navier-Stokes, they are non-trivial even in $d=1$.

A section-wise breakup : In section II we provide a derivation of the equations of motion
for the driven Heisenberg magnets (DHM) both from general symmetry arguments and by
analyzing explicit magnetized lattice-gas and continuum models. We next analize the steady
state phase diagram of the continuum equations of motion (Section III), and show that the
`high drive-temperature' steady state of the model is paramagnetic (Sect. IIIa). There
exists a nonequilibrium critical point whose critical  
exponents are determined using a dynamical renormalization group calculation (Sect. IIIb).
Lastly (Sect. IIIc) we study the `low drive-temperature' phase of the system which exhibits
spatio-temporal chaos. This chaotic phase may be `controlled' giving rise to a steady state
configuration with broken chiral symmetry (Section IV).  We end with a discussion of
results and future directions (Section V). The details of the diagrammatic calculations are
relegated to an appendix (Appendix A).

\section{Dynamics of Driven Heisenberg Magnets}

Consider a system of Heisenberg ($O(3)$) spins situated either on a (hypercubic) lattice or
a continuum in $d$-dimensions. Imagine a current of mobile species moving along one spatial
direction (say ${\hat z}$), and interacting with the $O(3)$ spins. The spins interact with
their neighboring spins via an exchange interaction (or any other short-range ferromagnetic
interaction). If we treat the mobile species as being `faster' (in a way which we will make
precise) than the spins, then we may ask for the effective dynamics of the Heisenberg spins
themselves. We first derive the effective dynamics of the spins using general symmetry
arguments and conservation laws. While such arguments are robust in themselves, they may
not be realizable in a given physical setting. To counter this criticism we also construct
explicit microscopic realizations of the DHM.

\subsection{General Symmetry Arguments}

In this section we explain, how to construct, on general grounds of symmetry, the leading
terms in the equations of motion of a uniaxially driven Heisenberg magnet.

First let us remind ourselves of the equations of motion
for spins {\em at thermal equilibrium} at temperature $T$. 
The probability of spin configurations 
$\{\vs_i\}$ of a general nearest-neighbor Heisenberg chain 
with sites $i$ is $\propto \exp(-H/T)$, with an energy
function  
\be
\label{heis}
H = - \sum_i J_i \vs_i \cdot \vs_{i+1}, 
\ee
where $J_i$ is the ferromagnetic exchange coupling between $i$ and $i + 1$. 
A spin $\vs_i$ at $i$ precesses as $\dot{\vs}_i = \vs_i \times \vh_i$ where 
\be
\label{heq}
\vh_i = -{\partial H \over \partial \vs_i} = J_i\vs_{i+1} + J_{i-1}\vs_{i-1}
\ee
is the local molecular field. Replacing $J_i \to J(x)$ 
and $\vs_i \to \vs(x)$ in the continuum limit, yields \cite{radha} 
$\dot{\vs}(x) = J(x) \vs \times \partial^2_x \vs + (dJ/dx) \vs \times \partial_x \vs +
...$. For the physically reasonable case where $J$ varies {\em periodically} about a mean 
value $J_0$, this reduces for long wavelengths to $\dot{\vs}(x) = J_0 \vs \times
\partial^2_x \vs$, which is invariant under $x \to -x$, even if the 
$H$ is not. If we impose a macroscopic distinction between $x$ and $-x$ in 
the form of an (admittedly artificial) $J$ varying linearly with $x$, 
say $J = ax$ where $a$ is a constant, 
we see that a term $a \vs \times \partial_x \vs$ arises. However, 
the other term so generated, namely, $J(x) \vs \times \partial^2_x \vs$,  
has a coefficient which grows with $x$, presenting problems 
in the limit of infinite system size. The dynamics in either of these 
cases conserves $\sum_i\vs_i$ since it commutes with $H$.  
Note that we have so far insisted on a local field arising from an energy 
function as in (\ref{heq}). 

Let us now ask what the most general equation of motion for $\vs$ 
would be, if we relax the $x \to -x$ symmetry {\em and} no longer insist 
on an energy function. This would clearly be the appropriate approach if 
the system were carrying a steady drift current 
of some mobile species -- particles, vacancies, heat -- in, say, the
${\bf \hat{x}}$ direction, {\em retaining isotropy in spin space}. 
Let us not specify at this stage how these additional degrees of freedom 
couple to the spins. It is sufficient to note that, irrespective 
of the nature of such couplings, the spins will be in a nonequilibrium state 
so that their dynamics will not follow from an energy function, and must be 
constructed anew. 
 
If we average over the degrees of freedom directly 
associated with the current (the mobile species are `faster' than the spins), their effect 
should be simply to modify the equations for 
the $\vs_i$ by allowing terms forbidden at {\em thermal} equilibrium.
On general symmetry grounds the new terms will clearly involve an odd
number of spatial derivatives. To lowest order in gradients, there are only 
two such terms: 
$v\partial_x {\vs}$ and $\lambda {\vs} \times \partial_x {\vs}$.  
The first term, representing advection by a mean drift, 
may be eliminated by a Galilean transformation $x \to x +
vt$, $t \to t$, leaving only the second term to reflect the drive.  
It is clear that the second term 
simply represents the continuum limit of asymmetric exchange, i.e.,  
$\vh_i = J_+\vs_{i+1} + J_-\vs_{i-1}$ 
with $\lambda \propto J_+ - J_-$ proportional 
to the driving rate. 
Such a term would be ruled 
out \cite{MAMAZ} at thermal equilibrium only because the dynamics {\em had} 
to be generated by (\ref{heis}) and (\ref{heq}). 

If we start with the usual Ma-Mazenko \cite{MAMAZ} dynamics for $\vec{S}$, with spin-conservation 
built in to both the systematic and noise terms, and then add in our novel precession 
($\lambda$) term, then standard perturbation theory for the noise and propagator 
renormalization yields, already at one-loop order, non-conserving terms of the form 
$({\vs} \cdot {\vs})^{n} \, {\vs}$ in the limit of zero external 
wavenumber even if these are not put in at the outset.   
This is because the $\lambda$ term, while rotation-invariant in spin space, 
is not the divergence of a 
current \cite{KPZ,NOTE} 
and, thus, does not conserve total spin.     
A renormalization-group theory of the long-wavelength dynamics of such 
a driven system must allow from the start for such nonconserving terms. 

For a general dimension $d \equiv d_{\perp}+1$, with anisotropic driving 
along one direction (${\|}$) only, the above arguments yield,  
to leading orders in a
gradient expansion, the generalized 
Langevin equation (which possesses spatial $O(d-1)$ symmetry),
\begin{eqnarray}
\frac{\partial {\vs}}{\partial t} & =  & \bigg( r_{\|}\partial_{\|}^2 
+ r_{\perp}\nabla_{\perp}^2 \bigg) {\vs} - v
{\vs}-\frac{u}{6}({\vs}\cdot{\vs}){\vs}
-\lambda {\vs}\times \partial_{\|}{\vs} \nonumber \\
 &   &  +g_{\|}{\vs}\times \partial_{\|}^2{\vs}
+g_{\perp}{\vs}\times \nabla_{\perp}^2{\vs} + \vec{\eta}\, ,
\label{eq:dyeq}
\end{eqnarray}
where we have allowed explicitly for spatial anisotropy in all coefficients.
The Gaussian, zero-mean nonconserving
noise $\vec \eta$ satisfies  
\begin{equation}
\langle \eta_{\alpha}({\bf x},t) \eta_{\beta}({\bf x}',t')\rangle = 
2B\,\delta_{\alpha \beta}\,\delta^d({\bf x}-{\bf x}') \delta(t-t')\,.
\label{eq:noise}
\end{equation}

Note that the noise in our equations is a {\em non-conserving scalar} noise. 
It is vectorial only in internal (spin) space but is a  scalar under 
{\em spatial} rotations. The correlation function of such a noise, at 
zero wavenumber, cannot detect spatial anisotropy. If the noise were {\em conserving}, 
the covariance at wavevector ${\bf q} = (q_{||}, {\bf q}_{\perp})$ would in general be 
proportional to $q_{||}^2 + \alpha q_{\perp}^2$, for some constant $\alpha \neq 1$. 
Such additive, {\em conserving} but anisotropic contributions to the noise doubtless 
exist here as well, but are irrelevant at small wavenumber relative to the 
nonconserving noise.   

Having established the form of the coarse-grained equations of motion on 
general symmetry grounds,
we now offer explicit microscopic models within which the novel, 
nonequilibrium precession term could arise.

\subsection{Construction of Explicit Examples}

Recall that we are trying to create a driven state with directed 
spatial anisotropy (i.e., distinguishing, say, $x$ from $-x$)  
while retaining isotropy in the spin degrees of freedom. Clearly, this cannot 
be achieved by means of a steady imposed spin current, since that would 
pick out a spin direction. We suggest here some approaches to achieve this.  
(I) Impose a steady current in some other species through the material 
and to show that this leads, via symmetry-allowed couplings between this additional 
degree of freedom and the spins, to the nonequilibrium precession term. 
The dynamics for the additional species is {\em conserving}; there is, strictly, 
speaking, no timescale on which 
the additional variables can be treated as fast and eliminated to yield 
an effective equation of motion for the spins alone. We must therefore 
employ a driven variant of model D, in the language of Ref. \cite{HH}. 
The effective spin dynamics we actually use in our paper should,
however, apply in the limit of an infinitely large diffusivity for the 
additional species or if processes that violate the conservation law 
for the additional species can be made to intervene. 
(II) Introduce the driving as a fluctuating magnetic field, statistically 
isotropic in spin space, but with short-ranged correlations which 
distinguish $x$ from $-x$. In both these cases (and, it seems clear, in general) 
it turns out that the exchange couplings have to be dynamical, not constant, 
in order to generate 
the nonequilibrium precession term. 
We now examine each of these cases.  

Consider a 1-dimensional lattice, each site $i$ of which can be either
vacant ($n_i(t) = 0$) or occupied by one particle ($n_i(t) = 1$) at time $t$. 
Each particle has {\em an attached 
Heisenberg spin} and may hop to the nearest neighbor at the right (left), if vacant,  
with probability $p$ ($q$). 
The spin $\vs_i(t)$ at an occupied 
site $i$ is the spin of the occupying particle.  
The local field at site $i$ is 
$\vh_i(t) = J_{i-1, i}\vs_{i-1}(t) + J_{i+1, i}\vs_{i+1}(t)$ 
where the exchange coupling $J_{ij}(t)$ determines, at time $t$, 
the field at $j$ due to the spin at $i$. 
If we let $J_{ij}$ depend on the configuration of the ${n_i}$ as   
$J_{ij}(t) = n_i(t)n_j(t)[J_1 + (J_2-J_1)n_j(t-1)]$, with  
the ${n_i}$ governed by an asymmetric exclusion process (ASEP) 
\cite{SZIA,MARROBOOK} the driving nonlinearity in (\ref{eq:dyeq}) is generated 
naturally.    
Explicitly, $J_{i, i \pm 1}(t) = J_1$ or $J_2$ according as $n_i(t-1) = 1$ or $0$,  
since the exchange coupling between, say, $i-1$ and $i$ is 
operative at time $t$ only if $n_{i-1}(t) n_i(t) = 1$,  
Assume for simplicity that the particles can hop only to the right.
Then a configuration $111$ at sites $i-1, \, i, \, {\rm and} \, i+1$ at time 
$t$ was either already present at time $t-1$ or arose from $011$ (by a right hop from 
$i-2$). Thus, for $111$ configurations, $J_{i-1,i}$ will be a weighted average of $J_1$
and $J_2$, while $J_{i+1,i}$ will be $J_1$. In the continuum limit, and averaging over 
the particle dynamics, we will get the driving term in (\ref{eq:dyeq}), with  
$\lambda \propto p - q$. 

The assumption of fast dynamics for the ASEP variable is justified 
if we allow evaporation-deposition 
as well, thus annulling the conservation law for particles. 
The derivation in our paper for the effective asymmetric 
exchange felt by a spin at a given site remains unaltered by this non-conservation  
since, unlike the hopping, the evaporation-deposition affects in an unbiased manner the 
sites on either side of the spin in question. Of course, particle non-conservation 
induces spin non-conservation trivially in this case.  

More generally, let us construct a dynamical exchange coupling 
$J_{ij}(t)$ for a collection of spins in a host lattice which lacks invariance under 
$x \to -x$. Imagine that the system is in a nonequilibrium environment,
stationary, isotropic and translation-invariant in a statistical sense,  
in the form of a fluctuating magnetic field $\vh_i$ at every 
point $i$ of the lattice. Assume that $J_{ij}(t)$ 
has a piece 
$\Delta J_{ij}(t) \propto \vh_i(t) \cdot \vh_j(t - \tau)$ where $\tau$ is some timescale 
intrinsic to the material. Since the material lacks $x \to -x$ invariance, the 
average $\langle \Delta J_{ij} \rangle \propto  
\langle \vh_i(t) \cdot \vh_j(t - \tau)\rangle$ is in general nonsymmetric in $ij$ 
if the sites $i,j$ are separated in the $x$ direction. Microscopically, this could 
happen if there is some underlying dynamical process, for instance structural relaxation, 
determining the exchange couplings between sites $i$ and $j$. Note that we do not 
need explicitly to drive anything through the system; asymmetry under 
$x \to -x$ and a nonequilibrium noise should suffice to produce this ratchet-like 
effect. In addition, the precession of spins in this fluctuating field 
implies a term $\vs_i \times \vh_i$ which is not spin conserving. 

Curiously, a term with precisely the form of our nonequilibrium precession term appears 
in the {\em equilibrium} dynamics of the {\em staggered} magnetization in the isotropic 
{\em antiferromagnet}, as presented in the work of \cite{radha}. The consequences of 
such a term appear not to be very dramatic there, presumably because of the presence 
of other couplings to the ferromagnetic order parameter. In addition, equation 
(\ref{eq:dyeq}) with no noise and with all coefficients on the right-hand side 
except $\lambda$ set to zero is known in the literature as the Belavin-Polyakov 
equation \cite{BP,rrbopk} and has been widely studied for its soliton solutions.  

Moving to a completely different description, consider the dynamics of chiral 
self-propelled particles \cite{tomchiral} described by a polar vector ${\vec n}$, 
suspended in a fluid and drifting along the ${\hat z}$ direction. 
A transient variation of $\vec{n}$ along ${\hat z}$ must lead  
\cite{PURCELL} to an overall rotation about $\vec{n}$. Shifting 
away the effect of the mean drift this clearly yields a dynamics of the 
form
$\partial_t {\vec n} \sim {\vec n} \times \partial_z {\vec n} + \ldots \,\, $
which is once again our nonequilibrium precession term. The consequences 
of this for the dynamics of self-propelled chiral objects 
remain to be worked out. 

\subsection{Fokker Planck Equation and absence of an FDT}

Since in the Langevin dynamics of this driven Heisenberg magnet, both the nonconservative
noise and deterministic terms arise from the external drive, there is no obvious relation
(such as the fluctuation-dissipation theorem (FDT)) between the variance of the nonconservative
noise and dissipation \cite{CHAIKIN}. This may easily be seen by constructing the 
Fokker-Planck equation \cite{VAN}
for the  probability distribution of spins $P(\{{\vs}\},t)$
corresponding to the Langevin equation (\ref{eq:dyeq}),
\begin{equation}
\frac{\partial P}{\partial t} =  - \frac{\partial}{\partial S_{\alpha}}
\left[ A_{\alpha} P - B \frac{\partial P}{\partial S_{\alpha}}\right]
\end{equation}
where
\begin{eqnarray}
{\vec A}  &  =  & \bigg(r_{\|}\partial_{\|}^2 
+ r_{\perp}\nabla_{\perp}^2 \bigg) {\vs} - v {\vs}
- \frac{u}{6}({\vs}\cdot{\vs}){\vs}
-\lambda {\vs} \times \partial_{\|}{\vs}
\nonumber \\
 &   &  +g_{\|} {\vs} \times \partial_{\|}^2{\vs}
+g_{\perp} {\vs} \times \nabla_{\perp}^2{\vs}
\end{eqnarray}
Since the stationary probability distribution of the steady
state configurations need not be the equilibrium canonical distribution 
$\exp(-F[{\vs}]/T)$ at a temperature $T$ (where $F$ is the free-energy functional), there
is no direct relation between the 
$B$ and the other parameters that enter the Langevin equation.

\section{Nonequilibrium Steady States}

Having discussed several microscopic realizations of the coarse-grained continuum dynamics Eq.
(\ref{eq:dyeq}), we will proceed to establish a `nonequilibrium phase
diagram' of steady states obtained by analyzing the stationary solutions
($\partial_t {\vs} = 0$) of Eq. (\ref{eq:dyeq}). Let us specify the
parameters in our `phase diagram'. 

In the {\em equilibrium, isotropic} limit, $\lambda = u = v \equiv 0$, $r_{\|} = r_{\perp} 
\equiv r$, the noise strength (which is conserved in the absence of the drive) vanishes at zero
wavenumber, and Eq.\ (\ref{eq:dyeq}) 
has a critical  
point where the renormalized $r$ goes to zero. In the {\em driven state}, the dynamics and
noise are nonconserving; the critical point is at $v = 0$, which in general takes place on
a curve in the temperature/driving-force plane. As the drive is taken to zero
there should be a crossover from nonequilibrium to equilibrium critical behavior. 
Our primary interest is in the behavior at a given
nonzero driving rate, for which it suffices to vary the temperature-like parameter 
$v$ (which we shall refer to as `drive-temperature') in (\ref{eq:dyeq}), keeping the rest fixed (with
$r_{\|}, \, r_{\perp}, u > 0$). 
Note that the variance of the nonconserved noise
$B$ is another temperature-like parameter. Since the FDT is violated by the external drive,
these two temperature-like variables $v$ and $B$ are not related to each other. 
The nonequilibrium phase diagram will therefore be parametrized by the two parameters
$v$ and $B$. 

\subsection{Dynamics at High Drive-Temperatures ($v>0$)}

At high drive-temperatures the steady state, obtained by setting 
$\partial_t {\vs}=0$ is
paramagnetic, characterized by 
$\langle {\vs}({\bf x},t) \rangle =0$ ($\langle \cdot\cdot\cdot
\rangle$ denotes an average over the noise $\vec \eta$)
and correlators . The effect of the drive is to change the correlation
lengths $\xi_{\perp}=\sqrt{r_{\perp}/v}+O(\lambda^2/\sqrt{v})-O(\lambda^2)$ 
and $\xi_{\|} = \sqrt{r_{\|}/v}+O(\lambda^2/\sqrt{v})-O(\lambda^2)$. 
Thus for $v>0$ (the paramagnetic phase) all correlations clearly decay on finite lengthscales
$\sim 1 / \sqrt{v}$  
and time-scales $\sim 1/v$, and nonlinearities are irrelevant. 
We find that this paramagnetic state is
stable under dynamical perturbations. This can be seen by 
writing ${\vs}({\bf x},t) = \langle {\vs}({\bf x},t) \rangle + {\vec u}({\bf
x},t)$, where ${\vec u}$ is an arbitrary small perturbation.
The time evolution of ${\vec u}({\bf x},t)$ to linear 
order is given  by
\begin{eqnarray}
\frac{\partial {\vec u}}{\partial t}= \left(r_{\perp} \nabla^2_{\perp}
+r_{\|} \partial^2_{\|}-v\right){\vec u}+\vec{\eta} \, ,
\label{eq:parastb1}
\end{eqnarray}
which on Fourier transformation reads
\begin{eqnarray}
{\vec u}_{\bf k}(t)&=&{\vec u}_{\bf k}(0)\exp(-\gamma_{\bf k}t) \nonumber \\
& &+\int_{0}^{\infty} dt'\vec{\eta}_{\bf k}(t')\exp(-\gamma_{\bf k}(t-t'))\,
,
\label{eq:parastb2}
\end{eqnarray}
where
\begin{equation}
\gamma_{\bf k}=r_{\perp} k^2_{\perp}+r_{\|} k^2_{\|}+v \, .
\end{equation}
This arbitrary perturbation ${\vec u}_{\bf k}$ always decays to zero when
$v>0$. The `paramagnetic phase' is therefore linearly stable.

\subsection{Dynamics in the Critical Phase $(v=0)$: New Critical Behavior}

A description of the nature of correlations on the critical surface $v = 0$, requires the
machinery of the Dynamical Renormalization Group (DRG). While there are several clear
expositions of this technique \cite{HWA} devised for specific problems, we hope that some
readers may benefit from a pedagogical treatment of DRG applied to our anisotropic
driven Heisenberg dynamics.

Let us first drop all nonlinear terms from the equations of motion. In the critical region,
defined by $v=0$, the linear theory
is massless resulting in divergent long wavelength fluctuations,
as can be seen by explicitly calculating the correlation function 
$C(x,t)=\langle {\vs}({\bf x}+{\bf x}^{'},t+t^{'})\cdot{\vs}
({\bf x}^{'},t^{'})\rangle$ from
Eq. (\ref{eq:dyeq}) when $u=\lambda=0$. The correlation function
is easily recast as a scaling form,
\begin{eqnarray}
C(x,t)=\frac{B}{r_{\|}}x_{\|}^{2\chi}
F_0\bigg(\frac{r_{\|}t}{x_{\|}^z},\frac{x_{\perp}}{x_{\|}^{\zeta}}
\sqrt{\frac{r_{\|}}{r_{\perp}}}\bigg) \, ,
\label{eq:lincor}
\end{eqnarray} 
where the roughening exponent $\chi=1-d/2$, the growth exponent $z=2$,
the anisotropy exponent $\zeta=1$, and $F_0$ is an analytic function of its
arguments.

What is the nature of these divergent fluctuations in the presence of the
nonlinear terms? This can be addressed via a standard implementation of the
dynamical renormalization group
(DRG) \cite{HWA} based on a perturbative expansion in the {\it small} couplings
$\lambda$ and $u$,
about the linear theory.
The perturbative corrections to the
correlation function may be equivalently viewed as arising from
modifications (renormalization) of the parameters $r_{\|},
\, r_{\perp} $ and $B$.
Renormalizability guarantees that the correlation function 
$C(x,t)$ will retain a scaling form as in Eq. (\ref{eq:lincor})
with modified exponents $z$, $\zeta$, $\chi$ and a new scaling function
\begin{eqnarray}
C(x,t)=x_{\|}^{2\chi}
F\bigg(\frac{t}{x_{\|}^z},\frac{x_{\perp}}{x_{\|}^{\zeta}}\bigg) \, .
\label{eq:scalcor}
\end{eqnarray}  
This implies that the critical region, defined by (renormalized) $v^R=0$, still has
divergent long wavelength
fluctuations. In what follows we assume renormalizability, which we justify {\it post facto} 
to lowest order in perturbation.

The perturbation expansion would make sense only if the couplings $\lambda$ and $u$ are small. This is
ensured by the DRG procedure which involves a systematic expansion about the {\it upper critical dimension}
$d_c$. As
we shall see, the renormalized couplings $\lambda$ and $u$ will flow to small values, of order
$\epsilon = d_c - d$. The upper critical dimension may be obtained by a simple power-counting argument.
Rescale space, time and the order
parameter as : 
$x_{\|}=bx_{\|}',\,{\bf x}_{\perp}=b^{\zeta}{\bf x}_{\perp}',\, t=b^zt'$
and ${\vs}=b^{\chi} {\vs}'$, where $b > 1$ is an arbitrary parameter.
We may reinterpret the effect of such a
rescaling as a change in the parameters\,; thus the form of Eq.
(\ref{eq:dyeq}) will remain unchanged if we change the parameters to their
primed values $r'_{\|}=b^{z-2}r_{\|}$,
$r'_{\perp}=b^{z-2\zeta}r_{\perp}$, $B'=b^{z-2\chi-\zeta (d-1)-1}B$,
$u'=b^{4-d}u$, $\lambda^{'}=b^{\chi+z-1}\lambda$, $g_{\|}^{'}
=b^{\chi+z-2}g_{\|}$, and $g_{\perp}^{'}=b^{\chi+z-2}g_{\perp}$. 
Demanding that the linear equation (in the
absence of nonlinearities, $u=\lambda=g_{\|}=g_{\perp}=0$)
be scale invariant, automatically fixes $\chi=1-d/2$, $z=2$, $\zeta=1$,
consistent with Eq. (\ref{eq:lincor}).

Inserting the values of the exponents evaluated within the linear theory,
we find that the couplings $\lambda$ and $u$ are {\it relevant}, i.e.,
they grow under rescaling, for dimensions $d < d_c = 4$, the {\it upper
critical dimension} for these couplings. Thus nontrivial exponents are expected for $d < d_c =4$
in the presence of nonlinearities.
The mean-field estimates in the previous
paragraph are valid for $d \geq 4$.
The other two couplings
$g_{\|}$ and $g_{\perp}$ have an upper critical dimension $d_c^{g}=2$; they are
therefore {\it irrelevant} for $d$ near $4$, and will be set to zero in the analysis
that follows.

We now carry out the DRG
calculation to compute the corrections to the new exponents and the scaling
function due to the nonlinear terms using a field theoretic approach \cite{HWA,AMIT,ZIN,tHOOFT}. 
The DRG procedure involves, as usual, two steps ---

(i) Start with a large-wavenumber cutoff
$\Lambda$. Solve the equation of motion iteratively in an expansion in $u$ and $\lambda$
Average over modes with wavenumber $k$ in the shell $(\Lambda /b, \Lambda)$, with $b =
e^{\ell}$, to obtain
``intermediate'' renormalized parameters $r_{\perp}^I$, $r_{\|}^{I}$, $\lambda^I$ etc.
in the equations of motion. This averaging is carried out to leading order in $\lambda$
and $u$, justified {\it post facto} by the fact that these couplings flow to small
values, of order $\epsilon$, at the DRG fixed point in dimension $d = 4 - \epsilon$.
(ii) Rescale space, time, and fields as in the previous
paragraph to restore the original cutoff $\Lambda$. The parameters $r_{\perp}^I$ etc.
will then acquire rescaling factors as above. Define $r_{\perp}(\ell) \equiv b^{z - 2
\zeta}r_{\perp}^I $, and similarly define $\ell$-dependent versions of the other
parameters in the equation. This may be recast as differential recursion equations in the 
various couplings.
\\

\noindent
(i)\underline{{\em Perturbative Calculation}} : 
The bare (unrenormalized) propagator $G_0({\bf k},\omega)$ and the
correlator
$C_0({\bf k},\omega)$ are defined from the linear theory as
\begin{eqnarray}
G_0({\bf k},\omega)&=&\frac{1}{r_{\|}k_{\|}^2+r_{\perp}k_{\perp}^2+
v-i \omega} \, ,\nonumber \\
C_0({\bf k},\omega)&=&\langle {\vs}({\bf k},\omega)\cdot {\vs}(-{\bf k}
,-\omega)\rangle \nonumber \\
&=&\frac{2B}{(r_{\|}k_{\|}^2+r_{\perp}k_{\perp}^2+v)^2+\omega^2} \, .
\end{eqnarray}

We next calculate the corrections
to the correlation functions from the nonlinearities, 
perturbatively
in the couplings $u$ and $\lambda$. On Fourier transforming
Eq. (\ref{eq:dyeq}) we obtain
\begin{widetext}
\begin{eqnarray}
S_{\alpha}({\bf k},\omega)=G_0({\bf k},\omega)\eta_{\alpha}-
\frac{i\lambda}{2}G_0({\bf k},\omega)\int \d q \d\nu [q_{\|}-(k_{\|}-q_{\|})]
\epsilon_{\alpha \beta \delta}S_{\beta}({\bf q},\nu)S_{\delta}({\bf k}-{\bf q}
,\omega-\nu) \nonumber \\
- uF_{\alpha\beta\gamma\delta}G_0({\bf k},\omega)
\int \d q_1\d q_2\d \nu_1 \d \nu_2
S_{\beta}({\bf q}_1,\nu_1)S_{\gamma}({\bf q}_2,\nu_2)
S_{\beta}({\bf k}-{\bf q}_1-{\bf q}_2,\omega-\nu_1-\nu_2) \, ,\nonumber \\
\label{eq:recurs}
\end{eqnarray}
\end{widetext}
where the Fourier transform is defined as
\begin{eqnarray}
{\vs}({\bf x},t)=\int \d  k \d \omega {\vs}({\bf k},\omega) 
e^{-i\omega t+i{\bf k}\cdot{\bf x}} \, ,
\end{eqnarray}
with the measure $\d q = d^d q / (2\pi)^d, \, \d \nu =d
\nu / 2\pi$ and the range of integration 
$0\leq |q| \leq \infty ,
\, -\infty \leq \nu \leq \infty$. The
coefficient of the cubic term is given by $F_{\alpha\beta\gamma\delta}
=(1/3)(\delta_{\alpha\beta}\delta_{\gamma\delta}
+\delta_{\alpha\delta}\delta_{\beta\gamma}+\delta_{\alpha\gamma}\delta_{\beta
\delta})$.

\begin{figure}
\includegraphics[scale=0.65]{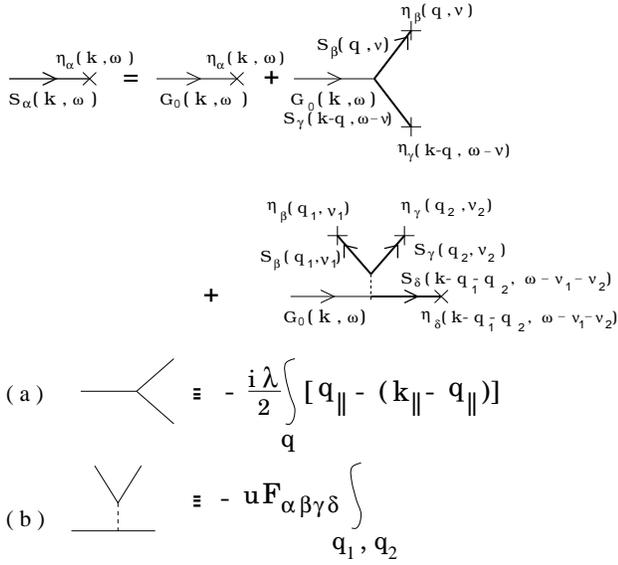}
\caption{\label{recurs}Graphs depicting the recursion relation in ${\vs}$. A thick line
with a cross at one end 
denotes ${\vs}({\bf k},\omega)$ and a thin arrowed line denotes $G_0({\bf k}
,\omega)$. The cross $\times$ denotes the noise $\eta_{\alpha}$.
(a) and (b) denote
the vertices coming from the $\lambda-$ and $u-$ couplings respectively.}
\end{figure}

Note that we have explicitly retained $v$ in the above expressions\,; after we have 
calculated the renormalized correlators, we shall set the {\it renormalized}  
$v^R=0$. This defines the critical phase.
To carry out the perturbative calculation effectively it is convenient to
rewrite the recursion relation Eq. (\ref{eq:recurs}) in terms of the
graphical representation displayed in
Fig.\ \ref{recurs}. The graphical representation is standard --- with $\rightarrow$ denoting 
the bare propagator $G_0({\bf k},\omega)$  and $\times$ denoting the noise $\eta_{\alpha}({\bf
k},\omega)$. The averaging over the noise is performed using
\begin{equation} 
\langle\eta_{\alpha}({\bf k}_1,\omega_1)\eta_{\beta}({\bf k}_2,\omega_2)
\rangle=2B \delta^d ({\bf k}_1+{\bf k}_2)\delta(\omega_1+\omega_2)
\delta_{\alpha\beta}\,.
\end{equation}

The renormalized propagator $G^R({\bf k},\omega)$
can be obtained perturbatively from Fig.\ \ref{recurs}. The lowest order (1-loop) correction
is given by,
\begin{widetext}
\begin{eqnarray}
 & G^R({\bf k},\omega)&=G_0({\bf k},\omega)-G_0({\bf k},\omega)^2 \bigg\{
\frac{\lambda^2}{4}
\int \d q\d\nu \bigg[(q_{\|}-(k_{\|}-q_{\|}))
(k_{\|}-(-q_{\|}))G_0({\bf k}-{\bf q},\omega-\nu) \nonumber \\
& & \times C_0({\bf q},\nu) +((k_{\|}-q_{\|})-q_{\|})(k_{\|}+(k_{\|}-q_{\|}))
G_0({\bf q},\nu)C_0({\bf k}-{\bf q},\omega-\nu)\bigg] \nonumber \\
&  &+\frac{5u}{3}\int \d q \d \nu C_0({\bf q},\nu) \bigg\}
\nonumber \\
& & =G_0({\bf k},\omega)+G_0({\bf k},\omega)^2\,\Sigma({\bf k},\omega) \, .
\label{eq:grec}
\end{eqnarray}
\end{widetext}
Note that the 1-loop renormalized propagator has terms to $O(\lambda^2)$ and $O(u)$ 
in the couplings. The self energy $\Sigma({\bf k},\omega)$ defined above contains all the 1-loop corrections
coming from the 
nonlinear terms.
The calculation of $\Sigma({\bf k},\omega)$ at the critical point
($v^R \rightarrow 0$) reduces to evaluating integrals --- these are singular in the infrared
$k \rightarrow 0$ 
limit for $d<4$. This apparent divergence reflects the relevance of nonlinearities below 4
dimensions; indeed it is this kind of divergence that the renormalization group procedure is geared to
handle. For now we note that
the integrals 
turn out to be logarithmically divergent,
both in the infrared and the ultraviolet $k\rightarrow \infty$ limit,
as $d\to4$; further the divergent pieces in both the limits turn out to be the
same \cite{AMIT}.
This allows us to use a procedure known as 
dimensional regularization \cite{tHOOFT}, to isolate the 
divergences as poles in $4-d\equiv\epsilon=0$. 
The ultraviolet divergences are controlled by a momentum cutoff $\Lambda$.
Details of the calculation
are
presented in Appendix A.

The 1-loop renormalized propagator $G^R$ may be written as
\begin{eqnarray}
\left[G^R({\bf k},\omega)\right]^{-1}&=&r^R_{\|}
k_{\|}^2+r^R_{\perp} 
k_{\perp}^2+v^R-i\omega \nonumber \\
&=&G_0^{-1}({\bf k},\omega)-\Sigma({\bf k},\omega) \, .
\label{eq:remginv}
\end{eqnarray}
to obtain the renormalized parameters $v^R$, $r^R_{\|}$
and $r^R_{\perp}$. Thus in order to read out the corrections to the bare
parameters, we need only compute the $\omega \to 0$ limit of $\Sigma$, and expand the
1-loop corrections to $\Sigma$ in powers of the external momenta, retaining only the coefficients
of $k^{0}$, $k_{\|}^2$ and $k_{\perp}^2$ (Appendix A).

The 1-loop corrections to the parameters are given by,
\begin{eqnarray}
r^R_{\|}=r_{\|}\bigg(1+\frac{\pi \tau v^{-\epsilon/2}}{4\epsilon}\bigg) \, ,
\label{eq:rpara1}
\end{eqnarray}

\begin{eqnarray}
{r}^R_{\perp}=r_{\perp}\bigg(1+\frac{5\pi \tau 
v^{-\epsilon/2}}{48\epsilon}\bigg) \, ,
\end{eqnarray}

\begin{eqnarray}
v^R=v-\pi^3\Lambda r_{\perp} \bigg(\frac{\tau}{8}+\frac{5\,\kappa
}{6}\bigg) \, ,
\end{eqnarray}
where the effective coupling constants $\tau$ and $\kappa$ are defined as
\begin{equation}
\tau=\frac{\lambda^2B}{2\pi^3(r_{\|}r_{\perp})^{3/2}} \, ,
\label{eq:tau}
\end{equation}
and
\begin{equation}
\kappa=\frac{uB}{2\pi^3r_{\|}^{1/2}r_{\perp}^{3/2}}\, .
\label{eq:kappa}
\end{equation}
As discussed the ultraviolet divergences have been regulated by an upper momentum cutoff $\Lambda$
while the infrared divergences appear as poles in
$\epsilon$ (dimensional regularization).

To evaluate the corrections to the other parameters we need to calculate the renormalized correlation
function 
$C^R({\bf k},\omega)=\langle {\vs}({\bf k},\omega)\cdot 
{\vs}(-{\bf k},-\omega)\rangle$ which to 1-loop 
satisfies
\begin{widetext}
\begin{eqnarray}
&C^R({\bf k},\omega)&=C_0({\bf k},\omega)
-\frac{\lambda^2}{4}
\int \d q \d \nu ((k_{\|}-q_{\|})-q_{\|})
(q_{\|}-(k_{\|}-q_{\|}))C_0({\bf k}-{\bf q},\omega-\nu)C_0({\bf q},\nu)
\nonumber \\
&  & \hspace{2 cm} \times G_0({\bf k},\omega)G_0(-{\bf k},-\omega) 
\nonumber \\
&  & = 2G_0({\bf k},\omega)G_0(-{\bf k},-\omega) B +
G_0({\bf k},\omega)G_0(-{\bf k},-\omega) \Sigma_B \, .
\end{eqnarray}
\end{widetext}
The function $\Sigma_B$ contains all the corrections coming from the
nonlinear couplings and is calculated in Appendix A. The
renormalization of the noise may now be easily determined via the   
definition $C^R({\bf k},\omega)
=2B^R G^R({\bf k},\omega)G^R(-{\bf k},-\omega)$,
\begin{eqnarray}
B^{R}=B\bigg(1+\frac{\pi \tau v^{-\epsilon/2}}{32\epsilon}\bigg)\, .
\end{eqnarray}

We have also computed the lowest order corrections to the
nonlinear couplings
(vertex corrections) defined as $-i \lambda^R =-i \lambda +
2\,\Gamma_{\lambda}$, and $-u^R=-u+\Gamma_u$, where the vertex functions
$\Gamma_{\lambda}$ and $\Gamma_{u}$ have been computed in Appendix A.
To 1-loop we find
\begin{equation}
\lambda^{R}=
\lambda\bigg(1-\frac{3\pi \tau v^{-\epsilon/2}}{128\epsilon}\bigg)\, ,
\label{eq:relambda}
\end{equation}
and 
\begin{equation}
u^R=u-(r_{\|}^{1/2}r_{\perp}^{3/2})
\bigg(\frac{11\pi^4\kappa^2}{12B\epsilon}-\frac{27\pi^4\tau^2}{8B\epsilon}
\bigg)\, .
\label{eq:reu}
\end{equation}

\noindent
(ii)\underline{\em Recursion Relations} : 
The relation between the renormalized and bare couplings allows one to write down the
scaling behavior of these couplings. First note that the parameter $v$ appears in the
propagator as a {\it mass} term; infrared divergences characterizing critical behavior
arise only in the massless theory. Trading this mass scale for a scale of length, we may
define an ``observation scale'' $\xi \equiv b a_0 \approx b/\Lambda = {v}^{-1/2}$, where
$a_0$ is a
microscopic cutoff length and $b>1$ is a pure number. Thus, large $k/v^{1/2}$ means small
$x/\xi$ which defines the critical region.

Of course if the limit $b \to \infty$ is taken straightaway, the renormalized couplings would diverge. 
The strategy is therefore to increase $b$ gradually, averaging the couplings over a small interval
between $b$ and $b + db$ --- this amounts to writing a differential equation for the evolution of the
couplings as the scale parameter $b$ changes.

In the critical region, one may define dimensionless couplings using the above observation scale; for
instance,
$\tilde{r}^R_{\|}(b)=r^R_{\|}(b\,a_0)^{z-2}$. As may easily be seen
from Eq. (\ref{eq:rpara1}), this dimensionless coupling satisfies,
\begin{equation}
\tilde{r}^R_{\|}(b)=
r_{\|}(b\,a_0)^{z-2}\bigg(1+\frac{\pi \tau
(b\,a_0)^{\epsilon}}{4\epsilon}\bigg)\, .
\label{eq:drpara}
\end{equation}
Applying the rescaling operator
$b{\partial}/{\partial b}$, we obtain,
\begin{eqnarray}
b\frac{\partial \tilde{r}^R_{\|}}{\partial b} = r_{\|}(b\,a_0)^{z-2}
\bigg(z-2+\frac{\pi \tau (b\,a_0)^{\epsilon}}{4}\bigg)\, .
\label{eq:drpara1}
\end{eqnarray}
Similarly define a dimensionless parameter for $\tau$;
$\tilde{\tau}^R(b)=\tau^R(b\,a_0)^{\epsilon}$.
Recalling that $\tau$ is of $O(\lambda^2)$ (Eq. (\ref{eq:tau})), we may
to lowest order in $\tau$ replace $r_{\|}$ and $\tau$ in
Eq. (\ref{eq:drpara1}) by their
renormalized values. 
Finally expressing in terms of $\ell = \ln b$, 
we arrive at the 1-loop differential recursion relations (the ultraviolet cutoff $\Lambda$
has been set to 1),
\begin{eqnarray}
\frac {\partial {\tilde {r}}^R_{\|}}{\partial \ell} & = & {\tilde {r}}^R_{\|} (
z-2+\frac{\pi}{4} {\tilde {\tau}}^R), \nonumber \\
\frac{\partial {\tilde {r}}^R_{\perp}}{\partial \ell}  
& =  & {\tilde {r}}^R_{\perp} (
z-2\zeta+\frac{5\pi}{48}{\tilde {\tau}}^R ), \nonumber \\
\frac{\partial {\tilde {B}}^R}{\partial \ell} & = & {\tilde {B}}^R \left[
z-2\chi-\zeta(d-1)-1+\frac{\pi}{32} {\tilde {\tau}}^R \right],
\nonumber \\
\frac{\partial {\tilde {\tau}}^R}{\partial \ell} & = &
{\tilde {\tau}}^R\, (\epsilon \zeta-\frac{35}{64}\pi {\tilde {\tau}}^R), 
\nonumber \\
\frac{\partial {\tilde {\kappa}}^R}{\partial \ell} & = &
{\tilde {\kappa}}^R\,(
\zeta\epsilon- \frac{11}{24}\,\pi \zeta {\tilde {\kappa}}^R -
\frac{\pi}{2}{\tilde {\tau}}^R)+ \frac{27}{16} \pi \zeta\,({\tilde {\tau}}^R)^2. 
\end{eqnarray}
\\
Recall that the couplings $\lambda$ and $u$ enter the perturbation 
theory in the dimensionless combinations $\tau \equiv (1/2 \pi^3)\lambda^2\,B
/\sqrt{r_{\|}^3\,r_{\perp}^3}$
and $\kappa \equiv (1/2 \pi^3) u\,B /\sqrt{r_{\|} r_{\perp}^{3}}$. The reader would have
noted that we have set the
irrelevant couplings $g_{\|} = g_{\perp} = 0$ for $d = 4 - \epsilon$ at the 
critical point, $v^R = 0$. 
\\

\noindent
(iii)\underline{\em Fixed Points and RG flows} : 
We expect the scaling behavior Eq.\ (\ref{eq:scalcor}) to hold at the critical point
in the limit $\ell \to \infty$. Thus the parameters ${\tilde {r}}^R_{\|}$,
etc., should be scale-independent for $\ell \to \infty$.
This necessarily implies that $\frac{\partial {\tilde {r}}^R_{\|}}{\partial \ell}=0$ 
as $\ell \to \infty$ and so on. In other words, the critical behavior at $v^R=0$  is
given by the {\it fixed points} of the recursion equations derived above,
\begin{eqnarray}
\frac{\partial \tilde{\tau}^*}{\partial \ell}=
\frac{\partial \tilde{\kappa}^*}{\partial \ell}=
\frac{\partial \tilde{r_{\|}}^*}{\partial \ell}=
\frac{\partial \tilde{r_{\perp}}^*}{\partial \ell}=
\frac{\partial \tilde{\lambda}^*}{\partial \ell}=
\frac{\partial \tilde{B}^*}{\partial \ell}=0\, .
\end{eqnarray}

The above equations yield four fixed points; by making small deviations from these fixed
points along the {\it remaining} directions in parameter space we can determine their
stability
(to linear order). The exponents $z$, $\zeta$ and
$\chi$ may be evaluated (to $O(\epsilon)$) at the stable fixed points :

\begin{enumerate}

\item[(A)] {\underline {$\tilde{\tau}^*=\tilde{\kappa}^*=0$}} is the `gaussian fixed point'
and is
stable for
$d>4$ and unstable for $d<4$. The exponents take their `mean field'
values $z=2$, $\zeta = 1$, and $\chi = 1-d/2$ at this fixed point.

\item[(B)] {\underline {$\tilde{\tau}^*=0$, $\tilde{\kappa}^*= 24\epsilon/11\pi$}} is
an unstable fixed point both for $d>4$ and $d<4$. 

\item[(C)] {\underline {$\tilde{\tau}^*=64\epsilon/35\pi$, $\tilde{\kappa}^*=
24\epsilon[3+\sqrt{12235}]/385\pi$}} is another
unstable fixed point both for $d>4$ and $d<4$. 

\item[(D)] {\underline {$\tilde{\tau}^*=64\epsilon/(35\pi)$, $\tilde{\kappa}^*=
36\epsilon\,[1+\sqrt{1409}]/385\pi$}} is a nontrivial `driven fixed
point' and is unstable
for $d>4$ but {\it stable} for $d<4$. Exponents take nontrivial values to
$O(\epsilon)$,
$z=2-16\epsilon/35$, $\zeta = 1-2\epsilon/15$ (anisotropic), and $\chi =
1-d/2$.
Note that $\chi$ does not change from its mean field value to this
order, since the quartic vertex $u$ plays no role at lowest order in $\epsilon$.
\end{enumerate}

\begin{figure}[h]
\includegraphics[scale=0.7]{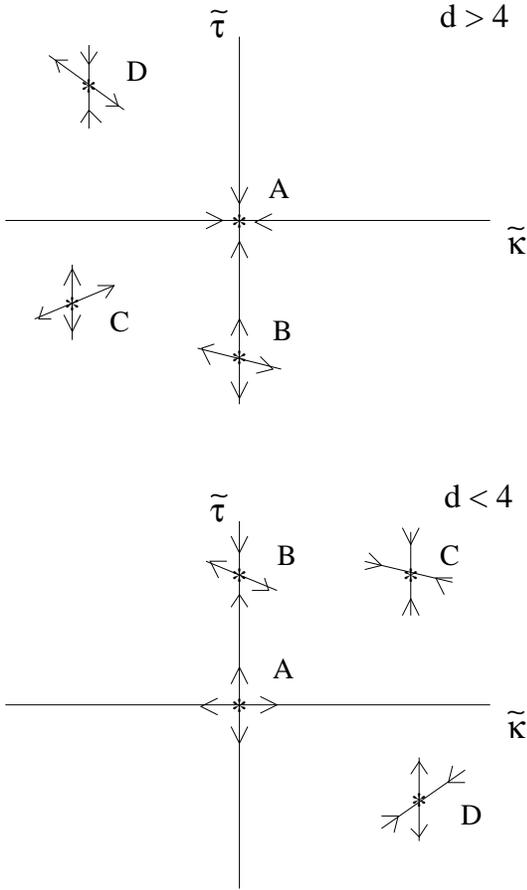}
\caption{\label{flow}Plot show fixed points and RG flows. For $d>4$, $A$ 
(Gaussian fixed point) is the only stable fixed point, while for $d<4$
there is a nontrivial stable driven fixed point $D$.}   
\end{figure}

Thus for $\epsilon = 4 - d > 0$, the nontrivial stable fixed point (D)  is associated with
the critical exponents $z=2-16\epsilon/35$, $\zeta = 1-2\epsilon/15$ and $\chi = 1-d/2$, to
lowest order in $\epsilon$. These exponents clearly place this critical behavior in a new
universality class and different from the anisotropic KPZ \cite{HWKAD}
In Fig.\ \ref{flow}, we exhibit the fixed points and the RG flow diagram to $O(\epsilon)$ in
a 2-dimensional subspace of the entire parameter space.

We end this section with a brief remark. There has been some recent discussion in the
literature \cite{UWE,UJZ} 
that for a particular class of driven $O(n)$ models, there exist no
stable fixed point to given order in $\epsilon$. The authors suggest, without proof, that
the RG flows in these models may be chaotic. We would like to stress that in our model,
there exists a stable fixed point in all dimensions, and so the critical behavior here,
atleast to $O(\epsilon)$, is not chaotic. However we will see in the next section that the
entire low drive-temperature phase, in the absence of noise, is chaotic. We will get back
to a discussion of these issues towards the end.

\section{Dynamics at Low Drive-Temperatures $(v<0)$: Spatio-Temporal Chaos}

We now investigate the low-temperature $v<0$ phase, {\em in the absence of noise}. 
It is convenient to
work  with dimensionless variables, obtained by  
rescaling $x_{\perp}$, $x_{\|}$, $t$ and ${\vs}$ in Eq.\
(\ref{eq:dyeq})\,; 
\begin{eqnarray*}
& &x_{\|} \rightarrow x_{\|}\sqrt{\frac{r_{\|}}{\vert v\vert}}, \, x_{\perp}
 \rightarrow x_{\perp}\sqrt{\frac{r_{\perp}}{\vert v\vert}}, \, t\rightarrow t \vert v\vert, \,
{\vs} \rightarrow {\vs}\sqrt{\frac{6\vert v\vert}{u}}, \nonumber \\
& & \, \vec{\eta} \rightarrow
\vec{\eta} \vert v\vert \sqrt{\frac{6\vert v\vert}{u}}, \, \lambda \rightarrow \lambda\sqrt{\frac{
6\vert v\vert}{r_{\|}}},  \, \\ 
& & g_{\|}\rightarrow g_{\|}\sqrt{\frac{6\vert v\vert}{ur^2_{\|}}},
\, g_{\perp}\rightarrow g_{\perp}\sqrt{\frac{6\vert v\vert}{ur^2_{\perp}}}, \, 
B\rightarrow B\frac{u\vert v\vert^{d/2-4}}{6r_{\|}^{1/2}r_{\perp}^{d/2-1}}\, ,
\end{eqnarray*}
leaving $\lambda$ as the only `drive' parameter 
in the equation of motion,
\begin{eqnarray}
\frac{\partial {\vs}}{\partial t}&=&\bigg(\partial_{\|}^2 
+\nabla_{\perp}^2 \bigg) {\vs}+
{\vs}-({\vs}\cdot{\vs}){\vs}
-\lambda {\vs}\times \partial_{\|}{\vs} \nonumber \\
& &+g_{\|}{\vs}\times \partial_{\|}^2{\vs}
+g_{\perp}{\vs}\times \nabla_{\perp}^2{\vs} \,\,.
\label{eq:sdyeq}
\end{eqnarray}

We first investigate the static, spatially {\it homogeneous} steady states :

(i) `Paramagnetic steady state' represented by $\langle S_{\alpha}
\rangle =0$ (average is taken over the steady state configurations) is a
solution of the stationary equations. It is easy to see from
Eq. (\ref{eq:parastb1}) that this steady state is linearly unstable.

(ii) `Ferromagnetic steady state' with broken O(3) symmetry 
represented by 
$\langle S_1 \rangle = \langle S_2\rangle = 0$ and $\langle
S_3\rangle = 1$ is also a solution of the stationary equations. 
This turns out to be linearly unstable too, as can be seen by perturbing
about this state by a small fluctuation ${\vec u}({\bf x},t)$
(to avoid a clutter of terms we set $g_{\perp}=g_{\|}=0$ since these
terms are of higher order in gradients than the $\lambda$ term),
\begin{eqnarray}
\partial_t u_1({\bf x},t)&=&\nabla^2 u_1({\bf x},t)+\lambda \partial_{\|}
u_2({\bf x},t) \, ,\nonumber \\
\partial_t u_2({\bf x},t)&=&\nabla^2 u_2({\bf x},t)-\lambda \partial_{\|}
u_1({\bf x},t) \, , \nonumber \\
\partial_t u_3({\bf x},t)&=&\nabla^2 u_3({\bf x},t)-2 u_3({\bf x},t)\, . 
\end{eqnarray}
Using the combination 
$u^{+}=u_1+i u_2$,
$u^{-}=u_1-i u_2$ and $u_3$ the above equations simplify in
Fourier space,
\begin{eqnarray}
\partial_t u^{+}_{\bf k}(t)&=&-k^2u^{+}_{\bf k}(t)+\lambda k_{\|}
u^{+}_{\bf k}(t) \, , \nonumber \\
\partial_t u^{-}_{\bf k}(t)&=&-k^2u^{-}_{\bf k}(t)-\lambda k_{\|}
u^{-}_{\bf k}(t) \, , \nonumber \\
\partial_t u_{3 \bf k}(t)&=&-k^2u_{3\bf k}(t)-2 u_{3 \bf k}(t)\, , 
\end{eqnarray}
clearly showing that $ u^{\pm}_{\bf k}(t)=u^{\pm}_{\bf k}(0)\exp(-k^{2} t \pm
\lambda k_{\|} t)$ are unstable at large wavelengths when
$k_{\|}<\lambda$.

(iii) We next look for static, spatially
inhomogeneous steady states; a natural candidate is the
 `helical steady state' which is more conveniently expressed in terms
of the variables $\rho \equiv \sqrt{S_1^2 + S_2^2}$ 
and $\phi \equiv \tan^{-1}(S_2/S_1)$. In these variables,   
Eq.\ (\ref{eq:dyeq}) for $g_{\|}=g_{\perp}=0$ becomes  
\begin{eqnarray}
\frac{\partial \rho}{\partial t}&=& \nabla^2 \rho -\rho (\nabla \phi)^2
+\rho -(\rho^2+S^2_3)\rho -\lambda\rho S_3 \partial_{\|} \phi \, ,
\nonumber \\
\frac{\partial \phi}{\partial t}&=& \nabla^2\phi+\frac{2}{\rho}({\bf \nabla}
\rho)\cdot({\bf \nabla}\phi)+\frac{\lambda}{\rho}(S_3\partial_{\|}\rho
-\rho\partial_{\|}S_3) \, , \nonumber \\
\frac{\partial S_3}{\partial t}&=& \nabla^2 S_3 + S_3 -(\rho^2+S^2_3)S_3
+\lambda \rho^2 \partial_{\|}\phi \, .
\label{eq:cleq}
\end{eqnarray}
A regular helix $\rho=a$, $\phi=px_{\|}$ and $S_3=b$
($a,b$ and $p$ are arbitrary constants) is a steady state 
solution
if the projection
of the local spins along the $\|$ axis $b$ and the pitch $1/p$ 
satisfy the following relations
\begin{equation}
2 b^2 = 1-a^2(1+\lambda^2)\pm
\sqrt{(a^2(\lambda^2+1)-1)^2-4a^4}\,,
\label{eq:defpro}
\end{equation}
and
\begin{equation}
2 p= -\lambda b \pm \sqrt{ \lambda^2b^2-4(R^2-1)}\, ,
\label{eq:pitch}
\end{equation}
where $R=\sqrt{a^2+b^2}$ is the magnitude of each spin.
The only free parameter $a$ is however bounded by 
$a<(3+\lambda^2)^{-1/2}$, coming from the requirement that
$b$ be real. 

\begin{figure}
\includegraphics[scale=0.55]{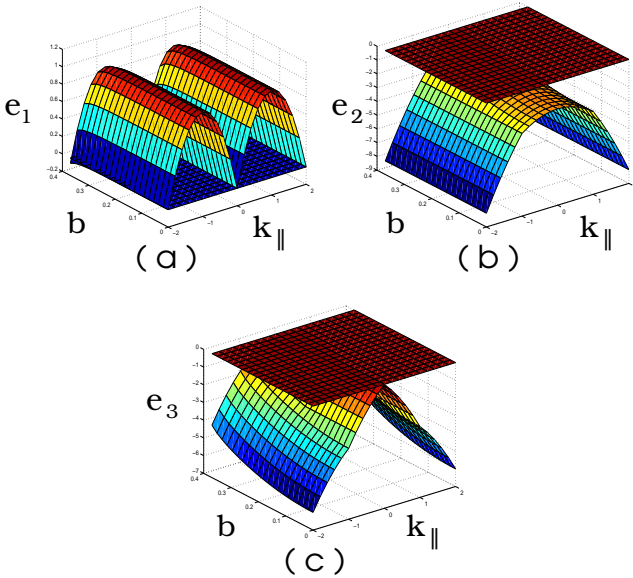}
\caption{\label{eigen}Plot of the real part of the eigenvalues $e_1$, $e_2$ and $e_3$
versus $a$ and $k_{\|}$ for $\lambda=2$.  The eigenvalue $e_1$ is positive
for all values of $a$ and $k_{\|}$ indicating linear instability. 
}   
\end{figure}

Unfortunately even the helical steady state is linearly unstable as we show
explicitly. Consider small
fluctuations about the helical steady state, 
$\rho=a+\delta\rho$, $\phi=px_{\|}+c+\delta \phi$ and $S_3=b+u$. To linear
order the
Fourier components of the fluctuations evolve as (for
simplicity we exhibit the modes with $k_{\perp}=0$)
\begin{widetext}
\begin{eqnarray}
\frac{\partial}{\partial t} \left(\begin{array}{c} \delta\rho_{\bf k} 
\\ \delta\phi_{\bf k} \\ u_{\bf k} \end{array} \right)&=& 
\underbrace{
\left( \begin{array}{ccc} -k^2_{\|}-2a^2 & -ik_{\|}a(2p+\lambda b) 
& -a(2b+\lambda p) \\
 ik_{\|}(2p+\lambda b)/a & -k^2_{\|} & -ik_{\|}\lambda \\
-2a(b-\lambda p) &  i\lambda k_{\|}a^2 &
 -k^2_{\|}+(1-a^2-3b^2) \end{array}\right) }_{D}
\left(\begin{array}{c} 
\delta\rho_{\bf k} 
\\ \delta\phi_{\bf k} \\ u_{\bf k} \end{array} \right) \, . \nonumber \\
\end{eqnarray}
\end{widetext}
The signature of instability is that the real part of any one of the
eigenvalues
of the matrix $D$ be positive. Fig.\ \ref{eigen} shows 2-dimensional
plots of the 
real part of the eigenvalues versus $a$ and $k_{\|}$
for a particular value of $\lambda$.
This shows that at least one eigenvalue has a positive real part for a
continuous band of $k_{\|}$. We have
checked that this
result holds for 
other values of $\lambda$. This implies that there is an infinity
of unstable spatially periodic steady states parametrized by $a$ (and for 
each value of $a$ there are two values of $b$ and $p$), a fact
that will be of some significance later.

The inhomogeneous helical steady state was suggested by the chiral nature
of the driving. It would be a hard task to do an exhaustive check of all 
inhomogeneous configurations for possible steady states.
Our strategy is therefore to solve the noiseless equations of motion
(\ref{eq:dyeq}) numerically starting from generic initial configurations
in both $1$ and $2$ dimensions.
The dynamics could either take the system to some other non-trivial
inhomogeneous stationary state or lead to temporally periodic or
chaotic configurations \cite{WOLFRAM}.

The numerical scheme for solving Eq. (\ref{eq:dyeq}) should be chosen carefully as the
linear derivative in the drive would give rise to numerical instabilities
if the standard Euler scheme of discretization were implemented
\cite{RECIPES}. We adopt an operator splitting method \cite{RECIPES} which allows
us to treat the dissipative terms and the drive separately 
under different discretization schemes. The dissipative part is 
solved using the standard
Euler method \cite{SR} and for the drive we use the following
algorithm. 
The time evolution of the spins with the drive alone is a
precession about the local magnetic field ${\vh}({\bf
x},t)=\partial_{\|}
{\vs}({\bf x},t)$. If ${\vh}({\bf x},t)$ were 
a constant in space and 
time, the local spin ${\vs}({\bf x},t)$ would have
precessed about this field, keeping
its magnitude $\vert {\vs}\vert$ fixed but changing 
its azimuthal angle $\phi$ 
(taking the direction of ${\vh}$ as the $z-$axis) by
$\vert{\vh}({\bf x},t)\vert\triangle t$ in a time interval of $\triangle t$. 
This would have been exact if the field ${\vh}$ were a constant\,; 
in our
case however ${\vh}({\bf x},t)$ depends on space and time and we
introduce
errors 
of $O(\triangle t)$. We choose
$\triangle  t$ small enough so as to reduce this error. The
advantage
of this method is that it does not
give rise to numerical instabilities and automatically preserves 
the magnitude of the local spin $\vert {\vs}\vert$ in time. 
In the simulation space and time are discretized with 
$\triangle x = 1$ and $\triangle t =0.0001$ on a
system of size $N=200$ (large enough to avoid finite size effects)
with periodic boundary conditions. The
local field is calculated  by the rule ${\vh}_i= ({\vs}_{i+1}
-{\vs}_{i-1})/\triangle x$. This field is used to update the local
spin by the precession algorithm.

\begin{figure}
\includegraphics[scale=1.2]{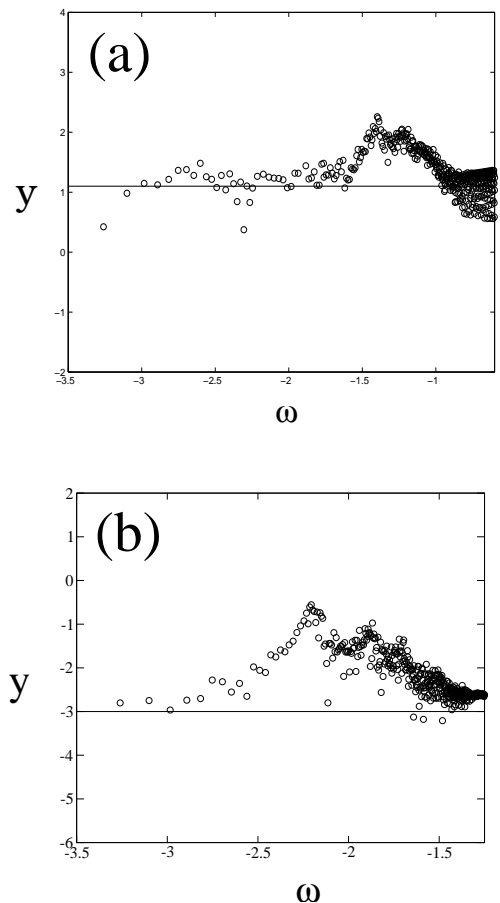}
\caption{\label{epow}
(a) Log-log plot of $y=\sqrt{\omega^2 \vert M_3(\omega)\vert^2}$ versus $\omega$
in $d=1$ showing the $1/\omega^2$ dependence of the power spectrum over
approximately $1.5$ decades. (b) Same plot for $d=2$.
}   
\end{figure}

Using this numerical scheme we can compute the time series of observables
like the magnetization and energy $E =  \int dx (\nabla
{\vs})^2 $. We first note that these quantities never seem to settle
to a constant value, strongly suggesting that no stable steady state
exists. The motion could therefore be either (quasi)periodic or
chaotic. This should be revealed in a power spectrum analysis\,; regular periodic 
motion would appear as a set of sharp delta functions. In Fig.\ \ref{epow}a we
display the power spectrum $P(\omega)=\vert M_3(\omega) \vert^2$ 
of the third component of the total
magnetization $M_3=\langle S_3 \rangle$ for data collected over more than
$3$ decades. The displayed spectrum has a clear smooth component with some small features which may be
erased by more averaging and more sophisticated binning.
In addition, we find that the power spectrum follows a power law ($1/\omega^2$)
behavior over roughly 2 decades. The power spectrum of the total energy
shows a similar behavior. These results strongly suggest that the
dynamics is temporally chaotic \cite{CROSS}. We have also checked that
this chaotic behavior shows up from a variety of initial configurations. The power spectrum
we obtain from a study of our model in 2 space dimensions shows similar features (Fig.\ \ref{epow}b).

\begin{figure}
\includegraphics[scale=1.0]{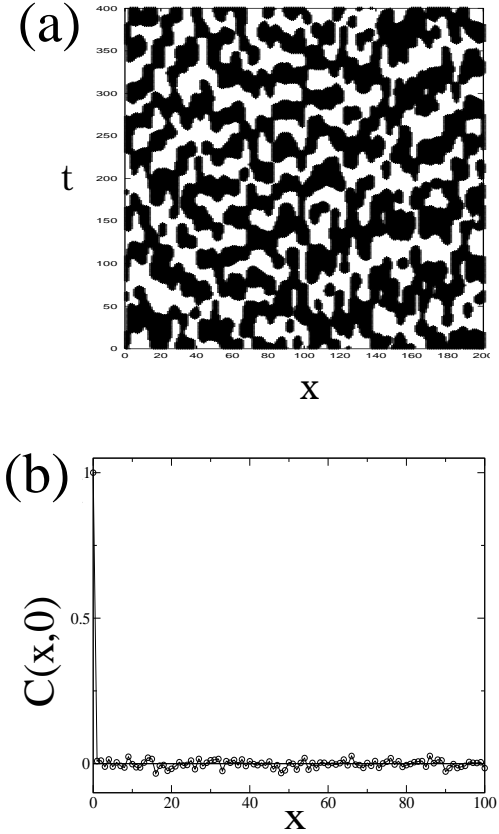}
\caption{\label{winphs}(a) Space-time plot of the signed local pitch, 
sgn$(p)\equiv$ sgn$(\partial_x \phi)$ (black and white patches denote
sgn$(p)=\pm 1$ respectively),
revealing spatio-temporal chaos.
(b)Plot of equal-time correlation function $C(x,0)$ vs $x$ (see text). }   
\end{figure}

Since the components of spin obey partial differential equations (PDEs),
we also check for spatial chaos. This is best visualized by constructing
space-time plots of local quantities. For instance, Fig.\ \ref{winphs}a is
a space-time plot of the signed local pitch, sgn$(p)\equiv$sgn$(\partial_x
\phi)$, suggesting spatio-temporal chaos\cite{CROSS}.

We also evaluate, in $d=1$, the space-time correlators of the spin 
$C(x,t)=\langle {\vs}({\bf x}+{\bf x}^{'},t+t^{'})\cdot{\vs}
({\bf x}^{'},t^{'})\rangle$ in this spatiotemporally chaotic phase. Fig.\ \ref{winphs}b shows that the equal-time spatial correlation function $C(x,0)$, decays exponentially with a correlation length of the order of the lattice spacing. On the other hand, the unequal-time correlator $C(0,t)$, seems to have a power-law over slightly more than a decade with an exponent $0.7$. While this is not inconsistent with spatio-temporal chaos \cite{CROSS}, one generally expects power law correlations in non-critical systems only if a conservation law or a Goldstone mode is present. We do not understand the origin of this power law decay at present,

Admittedly such  characterizations of spatio-temporal chaos
are only suggestive and should be made rigorous 
by studying the dependence of the
number of positive Lyapounov exponents on system
size. In spite of this,
we hope we have provided convincing evidence that the asymptotic
configurations in the low drive-temperature regime exhibit spatio-temporal
chaos. The numerical evidence we presented was for $d=1$ and $2$, and though we
cannot be sure whether this spatio-temporal chaos will persist at higher
spatial dimensions, we feel that this is quite likely. This is because in our
stability analysis of steady states done for arbitrary spatial dimension,
we failed to find any reasonable stable steady state configuration at low
drive-temperatures. Moreover, a Lyapunov stability analysis of the simpler
equation $\partial_t {\vs} = \lambda {\vs} \times \partial_{\|} {\vs}$
in arbitrary $d$ reveals that a tiny disturbance in the initial
conditions grows exponentially in time. Several questions arise, to which
we do not have answers at present, such as whether there exists a
low-dimensional chaotic attractor and if so what is its nature and
dimensionality.

The spatio-temporal chaotic phase that we just discovered has embedded in
it an infinity of unstable (spatially) periodic steady states. It seems likely \cite{OTT,SHIN}
that starting from generic initial conditions the
configuration of spins would eventually visit these periodic steady
states, although the time
taken to visit any one of these periodic configurations is unpredictable.
Since these periodic steady states are unstable, once visited, the
dynamics will veer the spin configurations away from it. 
We now ask whether we can arrange that the spin configuration stays put in
a prescribed periodic steady state having visited it ? This is the subject
of {\it control of spatiotemporally chaotic systems}, one of the most
important problems in modern chaos research \cite{OTT,SHIN}. There are
two aspects to the control of chaos, {\it stabilisation} and {\it
targeting}. Holding the periodic steady state having visited it is termed
stabilization. However since the time taken for this visit from an
arbitrary initial condition can be extremely large, it is desirable to
{\it target} a prescribed unstable periodic steady state. There have been
many proposals for controlling chaos in finite dimensional dynamical
systems \cite{OTT,SHIN}. However there has been very little work in the
more important area of control of spatio-temporal chaos in PDEs (which
correspond to an infinite dimensional dynamical system, see Ref.
\cite{SHIN} for a review). 
Accordingly, it is significant that we are able to stabilize, target, and hence
control spatiotemporal chaos in our model, as we now show.

\section{Targeting and Control: Emergence of Helical states}

\begin{figure}
\includegraphics[scale=0.23]{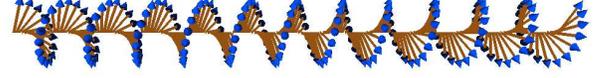}
\caption{\label{helix}Picture of the stabilized helical configuration from
numerical simulations.}   
\end{figure}

The helix solutions of Eq.  
(\ref{eq:cleq}) for $v < 0$ are an infinite family of unstable spatially periodic steady 
states of the type discussed in \cite{OTT}. 
These helical
steady states are parametrized by $a$, the projection of the spin along
the $\perp$ axis, $p$, the inverse pitch, and $b$, the projection of the spin
along the $\|$ axis.
Can chaos in our 
model be {\em controlled} so as to {\em stabilize} and {\it target} \cite{OTT} 
these helical states? Our 
control strategy focuses on the spin component $S_3$.
Thus for instance, in order to stabilize a specific helical configuration
(with fixed $a$, $b$ and $p$), we could in principle wait till the dynamics 
(presumably ergodic) 
eventually leads to this configuration, after which we apply small
perturbations to prevent $S_3$ from deviating from the value $b$. This
successfully {\it stabilizes} the prescribed helix, Fig.\ \ref{helix}.

\begin{figure}
\includegraphics[scale=0.45]{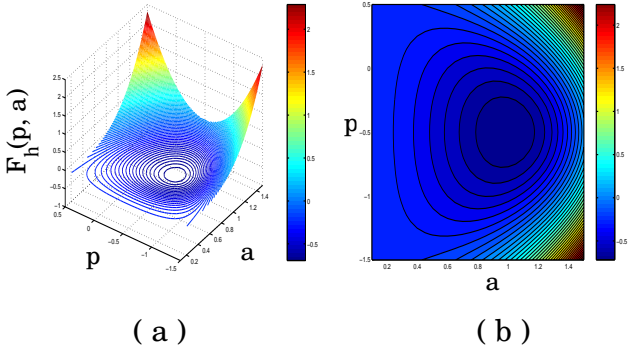}
\caption{\label{contour} (a) Contour plot of the `free-energy' $F_h$ as a function of $a$ and 
$p$. (b) Projection of plot (a) on the $pa$-plane. 
It is clear from the plots that there is a unique helical minimum.
}   
\end{figure}

In order to {\it target} this prescribed helix, we add to (\ref{eq:cleq}) 
terms which would arise from a uniaxial spin anisotropy energy $V_3 = r_3 (S_3^2 -
b^2)^2$ or $V_3 = r_3 (S_3-b)^2$. We find that a sufficiently large and positive $r_3$  
forces $S_3$ to take the value $b$ exponentially fast starting from
arbitrary initial configurations. The subsequent evolution,  
given by Eq. (\ref{eq:cleq}) on setting $S_3=b$ is (in $d$ dimensions),
\begin{eqnarray}
\frac{\partial \rho}{\partial t}&=& \nabla^2 \rho -\rho (\nabla \phi)^2
+\rho -(\rho^2+b^2)\rho -\lambda b\rho \partial_{\|} \phi \, ,
\nonumber \\
\frac{\partial \phi}{\partial t}&=& \nabla^2\phi+\frac{2}{\rho}({\bf \nabla}
\rho)\cdot({\bf \nabla}\phi)+\frac{\lambda b}{\rho}(\partial_{\|}\rho) 
 \, . 
\label{eq:spieq}
\end{eqnarray}
We now note that these equations can be recast as
purely relaxational dynamics,
\begin{eqnarray}
\frac{\partial \rho}{\partial t}&=&-\frac{\delta F}{\delta \rho} \, , \nonumber \\ 
\frac{\partial \phi}{\partial t}&=&-\frac{1}{\rho^2}
\frac{\delta F}{\delta \phi} \, , 
\end{eqnarray}
where the `free-energy functional' $F$ has the form of a {\it chiral
XY model} \cite{CHAIKIN}, 
\begin{eqnarray}
F&=&\frac{1}{2}\int d^dx \bigg[(\nabla \rho)^2+\rho^2(\nabla
\phi)^2-(\rho^2+b^2)\nonumber \\
& &+\frac{1}{2}(\rho^2+b^2)^2 
+\lambda b \rho^2 \partial_{\|} \phi \bigg] \,.
\label{eq:liap}
\end{eqnarray}
It is easy to see, using the chain-rule,  that
$F$ is a Lyapunov functional of the dynamics \cite{VAN},
\begin{eqnarray}
\frac{d F}{d t}&=&\int d^dx \left(\frac{\delta F}{\delta S_1}
\frac{\partial S_1}{\partial t}
+\frac{\delta F}{\delta S_2}\frac{\partial S_2}{\partial t}\right) \nonumber \\
&=&\int d^dx \left[-\bigg(\frac{\delta F}{\delta S_1}\bigg)^2
-\bigg(\frac{\delta F}{\delta S_2}\bigg)^2 \right] < 0 \, ,
\end{eqnarray}
which decreases monotonically in time.
Completing the squares, we see that $\partial_{\|} \phi$ appears in $F$ in the
combination $(1/2)\rho^2(\partial_{\|} \phi + \lambda b /2)^2$, which is minimized by 
the helix $\phi = - (1/2)\lambda b x_{\|}$. 
Hence starting from any initial configuration the system plummets towards
the minimum of this
$F$, which is a unique helix with parameters $b$, $a$ and $p$ ($p$ is
related to
$a$ and $b$ via Eq. (\ref{eq:pitch})). That there is a unique minimum
can also be seen by determining the `free-energy' $F_h$ of the helical
configurations from Eq. (\ref{eq:liap}) and plotting $F_h[p,a]$ against
$p$ and $a$ (Fig.\ \ref{contour}).

The stability of this `free-energy' minimizing helix can also be tested
directly from Eq.(\ref{eq:spieq}). As before, we perturb about this helix
: $\rho = a + {\tilde \rho}({\bf x},t)$,
and $\phi = px_{\|}+c+{\tilde \phi}({\bf x},t)$, and deduce the growth
of these perturbations to linear order,
\begin{eqnarray}
\frac{\partial \tilde \rho}{\partial t}&=& \nabla^2 \tilde \rho -2a^2
\tilde \rho 
-a(2p+\lambda b)\partial_{\|} \tilde \phi \, , \nonumber \\
\frac{\partial \tilde \phi}{\partial t}&=&\nabla^2\tilde \phi
+\frac{1}{a}(2p+\lambda b)
\partial_{\|}\tilde \rho \, .
\label{eq:lineq}
\end{eqnarray}
By going
to Fourier space we may evaluate the eigenvalues $\omega_1({\bf k})
$ and $\omega_2({\bf k})$ of the dynamical matrix, given by
\begin{eqnarray}
\omega_{1,2}({\bf k})=
-(k^2_{\|}+k^2_{\perp}+a^2) \pm \bigg\{ a^2+k^2_{\|}
(2p+\lambda b)^2\bigg\}^{1/2} \, .
\end{eqnarray}
The helix configuration is stable if these two eigenvalues have
negative real parts for all ${\bf k}$, which they do, as can be seen numerically by
substituting the values $a$ and $p$ take at the `free-energy' minimum
in the above expression.

Let us now see whether our control is robust against noise.
Indeed no discussion of the control of spatio-temporal chaos
is complete without considering the effects of
noise which might result in occasional escapes from the otherwise well
controlled system. Do these escapes lead to an instability of the targeted
configuration ? We therefore 
modify (\ref{eq:lineq}) by including the nonconservative 
Gaussian white noises  
$\eta_{\rho}$ and
$\eta_{\phi}$ to linear order.
We shall declare the controlled
helical state to be robust if the 
means $\langle \tilde \rho\rangle$ and $\langle \tilde 
\phi\rangle$ vanish and the variances are finite in the thermodynamic
limit. We therefore
ask for the statistics of small 
fluctuations with Fourier components $\tilde \rho_{\bf k}(t)$ and
$\tilde \phi_{\bf k}(t)$ about the controlled helical state, where 
$2 \pi/ L < k < \Lambda$ for a system of linear extent $L$ ($\Lambda$ is the
ultraviolet cutoff). 
It is clear from
(\ref{eq:lineq}) that the means $\langle \tilde \rho_{\bf k} \rangle$ and
$\langle
\tilde \phi_{\bf k}\rangle$ decay exponentially to zero: the relaxation time
for $\tilde \rho_{\bf k}$ is finite at small $k$, whereas that for 
$\tilde \phi_{\bf k}$ goes as $k^{-2}$.

To calculate the variances, 
note that the dynamics is governed in the mean by the 
Lyapounov functional (\ref{eq:liap}), and that the noise is spatiotemporally white. It 
follows \cite{CHAIKIN} that the steady-state configuration 
probability $P[\rho, \phi] \propto e^{- c F}$ where $c$ is an effective inverse
temperature. Since $F \simeq \int[ {\rm const} (\tilde \rho)^2 + {\rm const} 
(\nabla \tilde \phi)^2]$ for small fluctuations about the helical minimum, i.e., $P$ 
is approximately gaussian, this immediately implies that $\langle |\tilde \phi_{\bf k}|^2 \rangle \sim 
k^{-2}$ and $\langle |\tilde \rho_{\bf k}|^2 \rangle \sim {\rm const.}$ for small $k$. 
Thus the variance $\langle \tilde \rho^2 \rangle = 
\int_k \langle |\tilde \rho_{\bf k}|^2 \rangle$ 
($2\pi/L < k < \Lambda$),
is $L$-independent for $L \to \infty$ 
in any dimension $d$, whereas $\langle \tilde \phi^2 \rangle = 
\int_k \langle \vert\tilde \phi_{\bf k}\vert^2 \rangle$ diverges as $L$ and $\ln L$ respectively for $d = 1$
and
$2$, and is finite for $d > 2$. 

To obtain a more explicit asymptotic form for the the variance we calculate the 
equal-time correlation functions $C_{\tilde \rho}({\bf k},t)
=\langle \tilde \rho_{\bf k}(t)\tilde \rho_{-{\bf k}}(t)\rangle$ and 
 $C_{\tilde \phi}({\bf k},t)
=\langle \tilde \phi_{\bf k}(t)\tilde \phi_{-{\bf k}}(t)\rangle$ 
from the linearized equations of motion,
\begin{eqnarray}
\frac{\partial \tilde\rho}{\partial t}&=& \nabla^2 \tilde\rho
-2a^2\tilde \rho
-(2ap+\lambda ab)\partial_{\|} \tilde \phi 
+ \eta_{\rho} \, , \nonumber \\
\frac{\partial \tilde\phi}{\partial t}&=& \nabla^2\tilde\phi+
\frac{1}{a}(2p+\lambda b)\partial_{\|}\tilde \rho+\eta_{\phi} \, ,
\label{eq:linhelx}
\end{eqnarray}
where $\eta_{\rho}({\bf x},t)=\eta_1({\bf x},t)\cos(px_{\|})
+\eta_2({\bf x},t)\sin(px_{\|})$ and $\eta_{\phi}({\bf x},t)
=a^{-1}[\eta_2({\bf x},t)\cos(px_{\|})-\eta_1({\bf x},t)\sin(px_{\|})]$.
The noises $\eta_{\rho}({\bf k},t)$ and $\eta_{\phi}({\bf k},t)$
satisfy
\begin{eqnarray}
\langle \eta_{\rho}({\bf k},t) \eta_{\rho}({\bf k}',t')\rangle
&=&2B\delta_{{\bf k}
,-{\bf k}'} \delta (t-t') \, ,\nonumber \\
\langle \eta_{\phi}({\bf k},t) \eta_{\phi}({\bf k}',t')\rangle
&=&\frac{2B}{a^2}
\delta_{{\bf k},-{\bf k}'} \delta (t-t') \, , \nonumber \\
\langle \eta_{\rho}({\bf k},t) \eta_{\phi}({\bf k}',t')\rangle &= & 0 \, .
\end{eqnarray}
We may easily solve Eq. (\ref{eq:linhelx}) for Fourier transforms 
$\rho_{\bf k}$ and $\phi_{\bf k}$,
\begin{eqnarray}
\tilde\rho_{\bf k}(\omega)&=&\frac{1}{D({\bf k},\omega)}(\eta_{\rho}({\bf k},\omega)
(i\omega+k^2) \nonumber \\
& &-i\eta_{\phi}({\bf k},\omega)k_{\|}a(2p+\lambda b)) \, ,\nonumber \\
\tilde\phi_{\bf k}(\omega)&=&\frac{1}{D({\bf k},\omega)}(\eta_{\phi}({\bf k},\omega)
(i\omega+k^2+2a^2) \nonumber \\
& &-\frac{i}{a}\eta_{\rho}({\bf k},\omega)k_{\|}(2p+\lambda b)) \, ,
\label{eq:linsoln}
\end{eqnarray}
where 
\begin{equation}
D({\bf k},\omega)=-\omega^2+2i\omega(k^2+a^2)+k^2(k^2+2a^2)-k_{\|}^2(2p+
\lambda b)^2\, .
\end{equation}

We now compute the equal time correlation functions $C_{\tilde\rho}$ and 
$C_{\tilde\phi}$ averaged over the noise:
\begin{eqnarray}
C_{\tilde\rho}({\bf k})&=&\langle \tilde\rho_{\bf k}(t)\tilde\rho_{-{\bf
k}}(t)\rangle
=\int \d\omega\langle \tilde\rho_{\bf k}(\omega)\tilde\rho_{-{\bf
k}}(-\omega) \rangle
\nonumber \\
&=& B\frac{k^4+k_{\|}^2a^2(2p+\lambda b)^2}{2f_{+}f_{-}(f_{+}+f_{-})}
+\frac{B}{2(f_{+}+f_{-})}
\end{eqnarray}
where $f_{+}$ and $f_{-}$ are the roots of the equation $D({\bf k},\omega)
=0$,
\begin{eqnarray}
f_{\pm}=k^2+a^2\pm \bigg[a^4+k_{\|}^2(2p+\lambda b)^2 \bigg]^{1/2} \, .
\end{eqnarray}
and 
\begin{eqnarray}
C_{\tilde\phi}({\bf k})&=&\langle \tilde\phi_{\bf k}(t)\tilde\phi_{-{\bf
k}}(t)\rangle
=\int \d\omega\langle \tilde\phi_{\bf k}(\omega)\tilde\phi_{-{\bf
k}}(-\omega)\rangle
\nonumber \\
&=&\frac{B}{a^2}\frac{k^2+2a^2+k_{\|}^2(2p + \lambda b)^2}{2f_{+}f_{-}
(f_{+}+f_{-})} \nonumber \\
& &+\frac{B}{2a^2}\frac{1}{f_{+}+f_{-}} \, .
\end{eqnarray}

We use the $k\rightarrow 0$ behavior to evaluate the variances
$\Delta_{\tilde\rho}$ and $\Delta_{\tilde\phi}$. In this infrared limit 
\begin{eqnarray}
C_{\tilde \rho}({\bf k},t) &  \sim   &  \frac{B}{k^2+a^2} \, , \nonumber \\
C_{\tilde \phi}({\bf k},t) &  \sim   &  \frac{B}{k^2_{\perp}+k^2_{\|}
\bigg(1-\left(\frac{2p + \lambda b}{a}\right)^2\bigg)}\, .
\label{eq:gstone}
\end{eqnarray}
The variances $\Delta_{\tilde{\rho}}$ and 
$\Delta_{\tilde{\phi}}$
obtained on integrating the correlators over all ${\bf
k}$ and then taking the thermodynamic limit $L\to \infty$ depend
sensitively on the spatial dimension $d$. 
The variance of $\tilde \rho$ is finite in all dimensions,
\begin{equation}
\Delta_{\tilde{\rho}} = \left\{ \begin{array}{ll}
                 \pi/a & \mbox{$d=1\, \, ,$}
\\
                 \ln(1+(\Lambda/a)^2) & \mbox{$d=2 \, \, ,$}
\\
                 \mbox{finite}  & \mbox{$d=3 \, \, ,$}
                       \end{array}
                 \right.
\label{eq:rho}
\end{equation}
while the variance of $\tilde \phi$ diverges in $1$ and $2$ dimensions and
is finite in higher dimensions,
\begin{equation}
\Delta_{\tilde{\phi}} = \left\{ \begin{array}{ll}
                 L  &  \mbox{$d=1 \, \, ,$}
\\
                 \ln L  &  \mbox{$d=2\, \, ,$}
\\
                 \mbox{finite}  & \mbox{$d=3\, \, .$}
                       \end{array}
                 \right. 
\label{eq:phi}
\end{equation}

Thus {\it occasional excursions from the controlled state as a result of the noise 
do not lead to an instability of the targeted state for $d>2$}; the behavior for $d
\leq 2$ is no worse than for a thermal equilibrium $XY$ model.  

\section{Discussions and Future Directions}

In this paper we have constructed the simplest example of  a spatially
extended dynamics in which dissipation, (reversible) precession and spatially anisotropic driving
act in concert to produce spatio-temporal chaos in a whole region of
parameter space. The model, the classical Heisenberg magnet in $d$ space dimensions, is driven
by imposing a background steady current of heat or particles (or any other
{\it mobile} species) which couples to the Heisenberg spins. 
Our driven Heisenberg model (DHM)
is a natural generalization of the DDLG models to the case of a 
3-component {\em axial}-vector order parameter ${\vs}$, and is an important step in 
the exploration of dynamic universality classes \cite{HH} far from equilibrium
\cite{UWE, MRUIZ}. 

\begin{figure}
\includegraphics[scale=0.42]{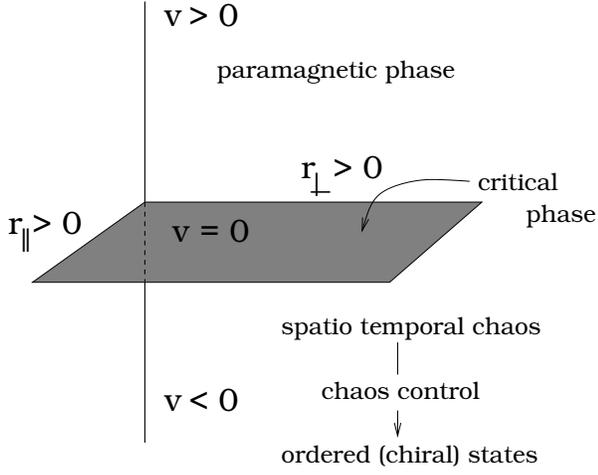}
\caption{\label{stdphs}
Non-equilibrium phase diagram as a function of the drive-temperature $v$. 
}   
\end{figure}

In the limit where the
mobile species is `fast',  the imposed steady current alters the effective dynamics
of the isotropic magnet so as to generate a driven precession term 
$\lambda {\vs}\times \partial_{\|}{\vs}$, that is responsible for all the remarkable phenomena we predict.
We summarize the nature of the asymptotic configurations in a 
`non-equilibrium phase diagram', Fig.\ \ref{stdphs}, as a function of 
the drive-temperature $v$.
The system exhibits a
`paramagnetic steady state' at high $v$ and a `critical steady
state' at $v=0$, with power-law correlations induced by
the driving even when the equilibrium Heisenberg magnet (without the
driving) is paramagnetic. The drive takes the system away from the
Wilson-Fisher fixed point leading to a new drive induced universality
class. At low drive-temperatures both the homogeneous and inhomogeneous
steady states are unstable. In particular, the system has an infinity of
spatially periodic unstable steady states which are helical. We have
provided evidence, at least in $d=1$ and $2$, that the dynamics at low
$v$ is spatio-temporally chaotic. We have found that the
spatio-temporal chaos may be `controlled' to target any desired helical
steady state. This control works even in the presence of noise in
dimensions $d>2$. The generic occurence of spatio-temporal chaos in our model is encouraging, given the intriguing
{\it formal} similarity to the nonlinearity $({\bf v}\cdot {\nabla}) {\bf v}$ in the Navier Stokes equation.

Our criterion for declaring the low temperature phase as spatio-temporally chaotic
is based on rather well established tests of power-spectrum analysis, space-time plots and decay of space-time correlators.
Admittedly such characterization is not very rigorous; ideally one would like
to analize the system size $L$ dependence of the Lyapounov
spectrum. 
We leave this for future investigation, as also several questions 
regarding
this spatio-temporal chaotic regime, such as the existence and
nature of a low dimensional attractor or the possibility of complex ordered states in higher dimensions,

Note that while the entire low drive-temperature phase is chaotic, the critical phase is not; the DRG analysis definitely shows  a stable fixed point in all dimensions. This is quite different from the possibility of a chaotic critical point, discussed in Ref. \cite{UWE}, for a general class of anisotropically driven $O(n)$ models along $d_{\parallel}$ directions. Their
driving appears as a nonequilibrium source of noise which is not bound by a fluctuation-dissipation relation. At the critical point of this model they find, to 1-loop, no stable fixed point when $d_{\parallel}>d_{\parallel}^*$. They tentatively put forward the possibility of spatio-temporal chaos at criticality. There have been earlier suggestions of 
{\it chaotic criticality} in equilibrium critical behavior, albeit in systems with quenched disorder
\cite{cardy}.
 
We conclude this paper by reminding the reader again of the variety of explicit lattice and continuum
realizations of our driven dynamics discussed in Sect.IIB  . We hope that this will stimulate 
a search for experimental systems, e.g., isotropic magnets carrying a 
steady particle or heat current, as well as model magnetized lattice-gas simulations, where
the predictions of our model can be tested. 

\begin{acknowledgements}
We thank A. Dhar, R.K.P. Zia, B. Schmittmann and 
U.C. T\"auber for discussions. JD gratefully acknowledges
the financial support provided by the Department of Energy (Materials 
Sciences Division of Lawrence Berkeley National Lab).
MR and SR thank the DST, India, for a Swarnajayanthi Grant and support
through the Centre for Condensed Matter Theory, respectively.
\end{acknowledgements}

\newpage

\appendix
\section{}

\begin{figure}
\includegraphics[scale=0.56]{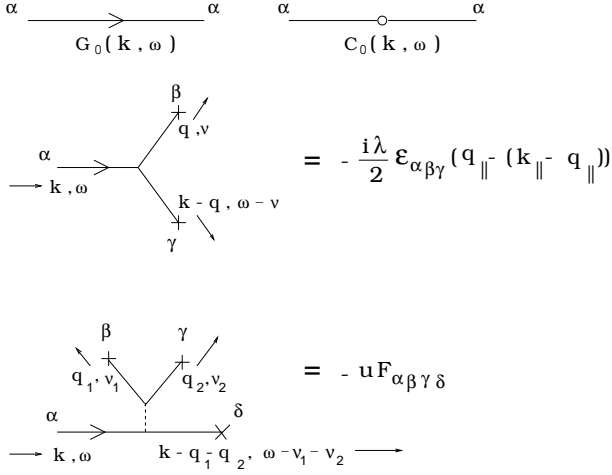}
\caption{\label{frule}Graphs for correlation functions and vertices at the tree level.
$F_{\alpha\beta\gamma\delta}=(1/3)[\delta_{\alpha\beta}\delta_{\gamma\delta}
+\delta_{\alpha\delta}\delta_{\beta\gamma}+\delta_{\alpha\gamma}\delta_{\beta
\delta}]$.
}   
\end{figure}

In this Appendix we present the details of perturbative calculation
mentioned in Sect. IV.
The diagrams corresponding to the lowest order terms in the
perturbation expansion are constructed following the rules shown in
Fig.\ \ref{frule}. 

\underline{\em Corrections to $G_0({\bf k},\omega)$}

\begin{figure}
\includegraphics[scale=0.55]{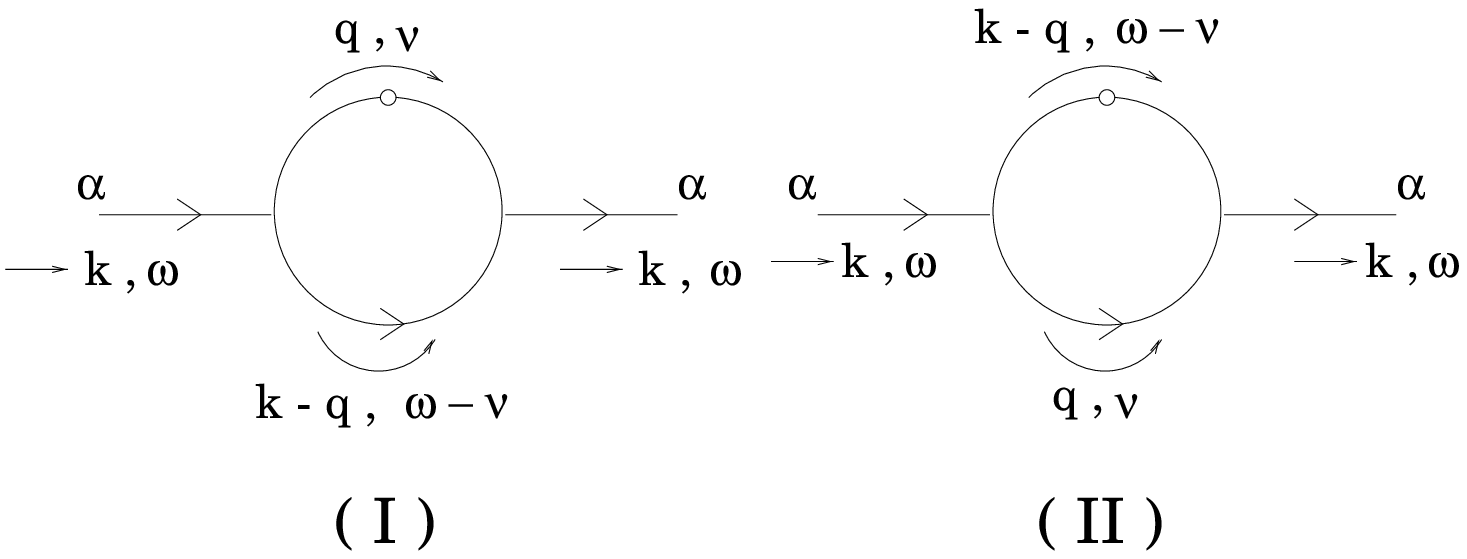}
\caption{\label{frule2pt}$O(\lambda^2)$ corrections to $G_0({\bf
k},\omega)$.
}   
\end{figure}

(I) Corrections from the $\lambda$ vertex :

Graphs (I) and (II) in Fig.\ \ref{frule2pt} show the one-loop
corrections
to $G_0({\bf k},\omega)$ due to the $\lambda$ vertex.
\begin{widetext}
\begin{eqnarray}
\Sigma_{\lambda}({\bf k},\omega)=-\frac{\lambda^2}{4}\int \d q \d\nu 
\bigg[(q_{\|}-(k_{\|}-q_{\|}))
(k_{\|}-(-q_{\|}))G_0({\bf k}-{\bf q},\omega-\nu)C_0({\bf q},\nu) \nonumber \\
+((k_{\|}-q_{\|})-q_{\|})(k_{\|}+(k_{\|}-q_{\|}))
G_0({\bf q},\nu)C_0({\bf k}-{\bf q},\omega-\nu)\bigg] \nonumber \\
=-\frac{2\lambda^2B}{4}\int \d q \bigg[\frac{(2q_{\|}-k_{\|})(k_{\|}+q_{\|})}
{\gamma^2({\bf q})}+\frac{(k_{\|}-2q_{\|})(2q_{\|}-q_{\|})}{\gamma^2({\bf k}-
{\bf q})}\bigg]\frac{1}{\gamma^2({\bf q})+\gamma^2({\bf k}-{\bf q})} \, .
\end{eqnarray}
\end{widetext}
In the above integrals
$\gamma^n({\bf k})$ is defined as
\[
\gamma^n({\bf k})=\bigg(r_{\|}k_{\|}^2+r_{\perp}
k_{\perp}^2+v\bigg)^n .
\]
We expand $\Sigma_{\lambda}({\bf k},0)$ in powers 
of $k_{\|}$ and $k_{\perp}$. Terms 
which are higher order than $O(k_{\|}^2)$ or $O(k_{\perp}^2)$ are 
ultraviolet convergent. Coefficients of the terms of order ${k}^0$,
$k_{\|}^2$ and $k_{\perp}^2$ denote changes to $v$, $r_{\|}$
and $r_{\perp}$ respectively.

\vspace{0.5cm}

(i) \underline{Term proportional to ${k}^0$}

This part of $\Sigma_{\lambda}({\bf k},0)$ does not have any 
infrared divergences but has
divergences in the ultraviolet. We introduce an upper momentum cutoff 
$\Lambda$
to get an estimate of the correction to $v$. We will denote the
self energy contribution coming from the $\lambda$ vertex to order
$k^0$ as $(\Sigma_0)_{\lambda}$,
\begin{eqnarray}
&(\Sigma_0)_{\lambda}& = -\frac{\lambda^2B}{2}\int 
\d q \frac{q_{\|}^2}{\gamma^4({\bf q})}
\nonumber \\
&  & = -\frac{\pi\Lambda\lambda^2B}{4r_{\|}^{3/2}r_{\perp}^{1/2}}\, .
\end{eqnarray}

(ii) \underline{Term proportional to $k_{\|}^2$}
 
\begin{eqnarray}
& (\Sigma_{\|})_{\lambda} = &-\frac{\lambda^2Bk_{\|}^2}{4}\int \d q 
\bigg[ \frac{1}{2\gamma^4({\bf q})}
+\frac{16 r_{\|}^2q_{\|}^4}{\gamma^8({\bf q})}-\frac{4 r_{\|} q_{\|}^2}
{\gamma^6({\bf q})}\bigg] \nonumber \\
&  &=\frac{\lambda^2 B v^{\epsilon/2} r_{\|}k_{\|}^2}
{8\pi^2\epsilon(r_{\|}r_{\perp})^{3/2}} \, ,
\end{eqnarray}
where
\begin{eqnarray}
\int \d q \frac{1}{\gamma^4({\bf q})}=\frac{r_{\|}v^{-\epsilon/2}}
{8\pi^2(r_{\|}r_{\perp})^{3/2}}\frac{1}{\epsilon} \, ,
\end{eqnarray}
\begin{eqnarray}
\int \d q \frac{q_{\|}^2}{\gamma^6({\bf q})}=\frac{v^{-\epsilon/2}}
{32\pi^2(r_{\|}r_{\perp})^{3/2}}\frac{1}{\epsilon} \, ,
\end{eqnarray}
\begin{eqnarray}
\int \d q \frac{q_{\|}^4}{\gamma^8({\bf q})}=\frac{r_{\|}v^{-\epsilon/2}}
{64\pi^2(r_{\|}^{5/2}r_{\perp}^{3/2})}\frac{1}{\epsilon} \, ,
\end{eqnarray}
have been evaluated using the standard integrals for the general form
$\int \d q (q_{\|}^a q_{\perp}^b)/\gamma^c({\bf q})$ given in
Ref. \cite{tHOOFT}. We have also had to make use of the asymptotic 
expansion for the Gamma function  
$\Gamma(-n+\epsilon)$, when $n$ is zero 
or any positive integer and $\epsilon \rightarrow 0$,
\begin{equation}
\Gamma(-n+\epsilon) = \frac{(-1)^n}{n!}\bigg[\frac{1}{\epsilon} + \bigg(1+
\frac{1}{2}+ \ldots + \frac{1}{n}-\gamma \bigg)+O(\epsilon) \bigg] \, ,
\end{equation}
where $\gamma$ is the Euler-Mascheroni constant
\cite{ABRAM}.

(iii) \underline{Term proportional to $k_{\perp}^2$} 

\begin{eqnarray}
& (\Sigma_{\perp})_{\lambda}=&-\frac{\lambda^2B k_{\perp}^2}{4}\int \d q q_{\|}^2 \bigg[
\frac{16 r_{\perp}q_{\perp}}{\gamma^8({\bf q})}-\frac{4r_{\perp}}{\gamma^6({\bf
q})}\bigg] \nonumber \\
& &=-\frac{5\lambda^2 B r_{\perp}k_{\perp}^2 v^{-\epsilon/2}}
{96 \pi^2 r_{\perp}(r_{\|}r_{\perp})^{3/2}}\frac{1}{\epsilon} \, ,
\end{eqnarray}
where we have evaluated the integral
\begin{eqnarray}
\int \d q \frac{q_{\|}^2q_{\perp}^2}{\gamma^8({\bf q})}=\frac
{v^{-\epsilon/2}}{48 \pi^2 
r_{\perp}(r_{\|}r_{\perp})^{3/2}}\frac{1}{\epsilon}\, .
\end{eqnarray}
\\
\\
(II) Corrections from the $u-$ vertex : \\
There is also a correction to the propagator $G({\bf k},\omega)$
coming from the $u-$ vertex\,; to $O(\epsilon)$ this is shown in the figure
below.

\begin{figure}
\includegraphics[scale=0.75]{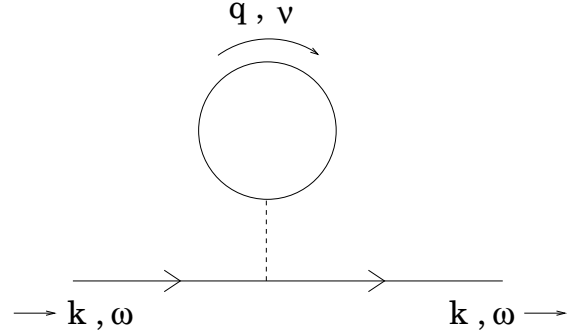}
\caption{$O(u^2)$ corrections to the ${\bf k}=0$ part of
$G_0({\bf k},\omega)$. 
}   
\end{figure}

To 1-loop, the correction to $G({\bf k},\omega)$ from this
interaction comes only from the ${\bf k}=0$ piece given by,
\begin{eqnarray}
& (\Sigma_0)_u & = -\frac{5u}{3}\int \d q \d \nu C_0({\bf q},\nu) 
\nonumber \\
&  & =-\frac{5uB}{3}\int \d q \frac{1}{\gamma^2({\bf q})} \nonumber \\
&  & = -\frac{5\pi\Lambda uB}{6 r_{\|}^{1/2}r_{\perp}^{3/2}}\, . 
\end{eqnarray}

The net 1-loop correction to the propagator $G$ adds up to 
\begin{equation}
\Sigma({\bf k},\omega)=[(\Sigma_0)_{\lambda}+(\Sigma_0)_u ] 
+(\Sigma_{\|})_{\lambda} k_{\|}^2+(\Sigma_{\perp})_{\lambda}
k_{\perp}^2\,.
\end{equation}

\underline{\em Corrections to $C_0({\bf k},\omega)$}
\begin{figure}
\includegraphics[scale=0.6]{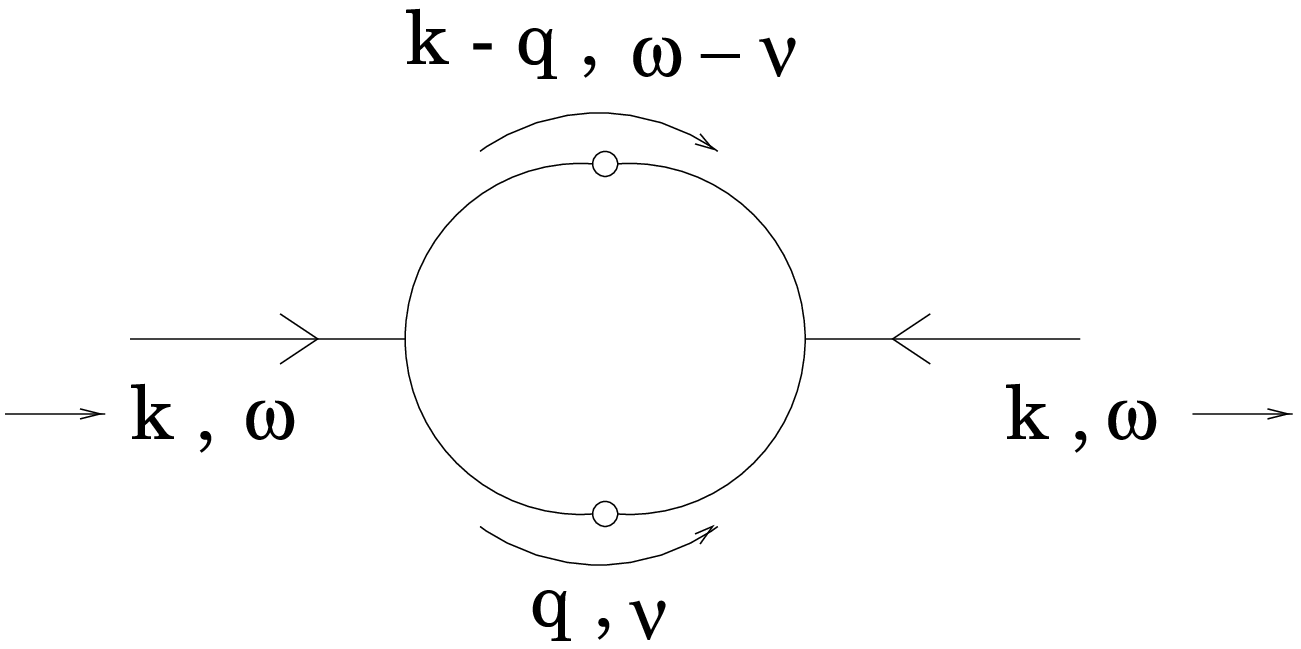}
\caption{\label{frule2ptc}$O(\lambda^2)$ corrections to $C_0({\bf k},\omega)$.
}   
\end{figure}
\begin{widetext}
\begin{eqnarray}
&\Sigma_{B}=& -\frac{\lambda^2}{4}\int \d q \d\nu ((k_{\|}-q_{\|})-q_{\|})
(q_{\|}-(k_{\|}-q_{\|}))C_0({\bf k}-{\bf q},\omega-\nu)C_0({\bf q},\nu)
\nonumber \\
&  &= \lambda^2B^2 \int \d q \frac{q_{\|}^2}{\gamma^6({\bf q})} \nonumber \\
&  &=\frac{\lambda^2B^2v^{-\epsilon/2}}{32\pi^2(r_{\|}r_{\perp})^{3/2}}
\frac{1}{\epsilon} \, .
\end{eqnarray}
\end{widetext}

The renormalized couplings are obtained from the vertex corrections to
$\lambda$ and $u$.

\underline{\em Corrections to the $\lambda-$ vertex} \\
The corrections to $\lambda$ come from the three-point correlators,

\begin{figure}
\includegraphics[scale=0.43]{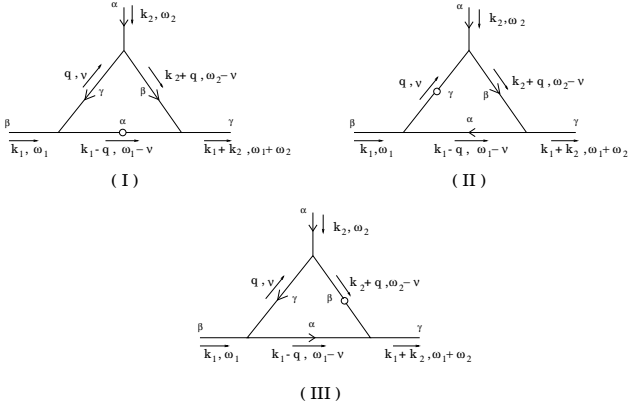}
\caption{$O(\lambda^3)$ correction to the $\lambda-$ vertex.
}   
\end{figure}
The contribution from (I) to the vertex function $\Gamma_{\lambda}$
is given by,
\begin{widetext}
\begin{eqnarray}
& & \bigg(\frac{-i\lambda}{2}\bigg)^3 \int \d q \d\nu \bigg[ ((k_{1\|}-q_{\|})
 +k_{1\|})((k_{2\|}+q_{\|})+q_{\|})((k_{1\|}+k_{2\|})+(k_{1\|}
-q_{\|}))\bigg] \times \nonumber \\
& &G_0({\bf q},\nu)G_0({\bf k}_2+{\bf q}, \omega_2-\nu)
C_0({\bf k}_1-{\bf q},\omega_1-\nu) \nonumber \\
& &=\frac{i\lambda^3B}{16}\int \d q \frac{(8k_{1\|}+k_{2\|})q_{\|}^2}{\gamma^6(
{\bf q})}\, .
\end{eqnarray}
\end{widetext}
Likewise the contribution from (II) to $\Gamma_{\lambda}$ is
\begin{widetext}
\begin{eqnarray}
& & \bigg(\frac{-i\lambda}{2}\bigg)^3 \int \d q \d\nu \bigg[(-k_{1\|}-q_{\|})
((k_{2\|}+q_{\|})+q_{\|})((k_{1\|}+k_{2\|})+(k_{1\|}
-q_{\|}))\bigg]\times \nonumber \\ 
& &G_0({\bf k}_1-{\bf q},\omega_1-\nu)G_0({\bf k}_2+{\bf q}, \omega_2-\nu)
C_0({\bf q},\nu) \nonumber \\
& &=-\frac{i\lambda^3B}{32}\int \d q \frac{(2k_{1\|}+k_{2\|})q_{\|}^2}{\gamma^6(
{\bf q})} \, ,
\end{eqnarray}
\end{widetext}
while (III) gives
\begin{widetext}
\begin{eqnarray}
& & \bigg(\frac{-i\lambda}{2}\bigg)^3 \int \d q \d\nu 
\bigg[ (-(k_{2\|}+q_{\|})
 -(k_{1\|}+k_{2\|}))((k_{2\|}+q_{\|})+q_{\|})((k_{1\|}-q_{\|})+k_{1\|})
\bigg]\times \nonumber \\ 
& &G_0({\bf q},\nu)G_0({\bf k}_1-{\bf q}, \omega_1+\nu)
C_0({\bf k}_2+{\bf q},\omega_1+\nu) \nonumber \\
& &=-\frac{i\lambda^3B}{32}\int \d q \frac{(2k_{1\|}-5k_{2\|})q_{\|}^2}
{\gamma^6({\bf q})} \, .
\end{eqnarray}
\end{widetext}
The three contributions combine to give
\begin{eqnarray}
&\Gamma_{\lambda}(2k_{1\|}+k_{2\|})=& \frac{3i\lambda^3B}{16}\int \d q 
\frac{q_{\|}^2(2k_{1\|}+k_{2\|})}
{\gamma^6({\bf q})} \nonumber \\
&  & =\frac{3i\lambda^3B(2k_{1\|}+k_{2\|})v^{-\epsilon/2}}{512 \pi^2
(r_{\|}r_{\perp})^{3/2}}\frac{1}{\epsilon} \, .
\end{eqnarray}

\underline{\em Corrections to the $u-$ vertex}\\
There are two contributions to the vertex function $\Gamma_u$
to $O(u^2)$ and $O(\lambda^4)$.

\begin{figure}
\includegraphics[scale=0.7]{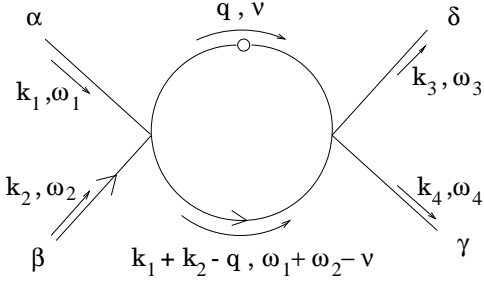}
\caption{$O(u^2)$ correction to the $u-$ vertex.
}   
\end{figure}
The $O(u^2)$ graph gives
\begin{eqnarray}
& &F_{\alpha\beta\gamma\delta}\frac{11u^2}{9}\int \d q\d\nu C_0({\bf q},\nu)G_0({\bf k}_1+{\bf k}_2
-{\bf q},\omega_1+\omega_2-\nu) \nonumber \\
=& &F_{\alpha\beta\gamma\delta}
\frac{11u^2B}{18}\int \d q \frac{1}{\gamma^4({\bf q})} \nonumber \\
=& &F_{\alpha\beta\gamma\delta}\frac{11u^2B}{48\pi^2
r_{\perp}^{3/2}r_{\|}^{1/2}}\frac{1}{\epsilon}\, .
\end{eqnarray}

\begin{figure}
\includegraphics[scale=0.7]{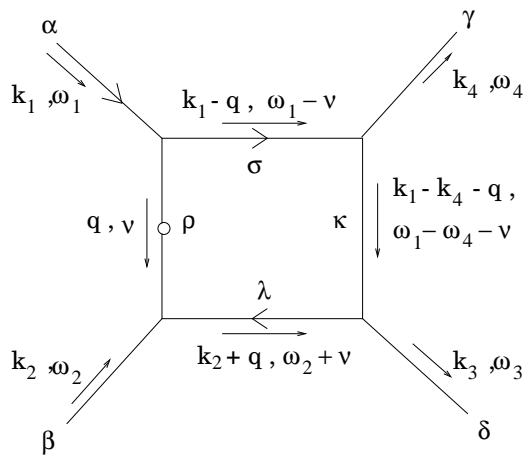}
\caption{$O(\lambda^4)$ correction to the $u-$ vertex.
}   
\end{figure}

The $O(\lambda^4)$ graph gives the following contribution :
\begin{widetext}
\begin{eqnarray}
& &-1152F_{\alpha\beta\gamma\delta}\bigg(\frac{\lambda}{4}\bigg)^4
\int \d q \d\nu ((k_{1\|}-q_{\|})-q_{\|})(k_{4\|}-(k_{1\|}-k_{4\|}-q_{\|}))
(k_{3\|}-(k_{2\|}+q_{\|}))(k_{2\|}+q_{\|})\times \nonumber \\
& &G_0({\bf k}_1-{\bf q},\omega-\nu)
G_0({\bf k}_1-{\bf k}_4-{\bf q},\omega_1-\omega_2-\nu)G_0({\bf k}_2+{\bf q},
\omega_2+\nu)C_0({\bf q},\nu) \nonumber \\
& & -1152F_{\alpha\beta\gamma\delta}\bigg(\frac{3\lambda^4B}{4^3}\bigg)
\int \d q \frac{q_{\|}^2}{\gamma^8({\bf q})} \nonumber \\
& & =-F_{\alpha\beta\gamma\delta}\frac{27\lambda^4}{32\pi^2 
r_{\perp}^{3/2}r_{\|}^{1/2}\epsilon} \, .
\end{eqnarray}
\end{widetext}
The net correction to the $u-$ vertex is,
\begin{eqnarray}
&\Gamma_{u} &=F_{\alpha\beta\gamma\delta}\frac{11u^2B}{48\pi^2
r_{\perp}^{3/2}r_{\|}^{1/2}}\frac{1}{\epsilon}-
F_{\alpha\beta\gamma\delta}\frac{27\lambda^4}{32\pi^2 
r_{\perp}^{3/2}r_{\|}^{1/2}\epsilon} \nonumber \\
&  & =F_{\alpha\beta\gamma\delta}(r_{\|}^{1/2}r_{\perp}^{3/2})
\bigg[\frac{11\pi^4\kappa^2}{12B\epsilon}-\frac{27\pi^4\tau^2}{8B\epsilon}
\bigg]\, ,
\end{eqnarray}
where $\kappa$ and $\tau$ are defined by
\begin{equation}
\kappa=\frac{u B}{2\pi^3(r_{\|}^{1/2}r_{\perp}^{3/2})} \, ,
\end{equation}
\begin{equation}
\tau=\frac{\lambda B}{2\pi^3(r_{\|}^{3/2}r_{\perp}^{3/2})}\, .
\end{equation}

\bibliography{drift}

\end{document}